\documentclass[journal]{IEEEtran}
\usepackage{amssymb}
\usepackage{amsbsy}
\usepackage{amsmath}
\usepackage{amsthm} 
\usepackage{setspace}
\usepackage{epsfig}
\usepackage{cite}
\usepackage{graphics}
\usepackage{epstopdf}
\usepackage{mathrsfs}
\usepackage[above]{placeins}
\usepackage{algorithm}
\usepackage{algpseudocode}
\usepackage{multirow}
\usepackage{algorithmicx}
\usepackage{tikz}
\usepackage{stfloats}

\begin{document}

\title{On the Geometric Ergodicity of\\
 Metropolis-Hastings Algorithms  \\
for Lattice Gaussian Sampling}

\author{Zheng~Wang,~\IEEEmembership{Member, IEEE,}
        and Cong~Ling,~\IEEEmembership{Member, IEEE}
\thanks{This work was presented in part at the IEEE International Symposium on Information Theory (ISIT), Hong Kong, China,
July 2015.

Z. Wang and C. Ling are with the Department of Electrical and Electronic Engineering, Imperial
College London, London SW7 2AZ, United Kingdom (e-mail: z.wang@ieee.org, cling@ieee.org).

This work was supported in part by FP7 project PHYLAWS (EU FP7-ICT 317562).

}}

\maketitle

\begin{abstract}
Sampling from the lattice Gaussian distribution has emerged as an important problem in coding, decoding and cryptography. In this paper, the classic Metropolis-Hastings (MH) algorithm in Markov chain Monte Carlo (MCMC) methods is adopted for lattice Gaussian sampling. Two MH-based algorithms are proposed, which overcome the limitation of Klein's algorithm. The first one, referred to as the independent Metropolis-Hastings-Klein (MHK) algorithm, establishes a Markov chain via an independent proposal distribution. We show that the Markov chain arising from this independent MHK algorithm is uniformly ergodic, namely, it converges to the stationary distribution exponentially fast regardless of the initial state. Moreover, the rate of convergence is analyzed in terms of the theta series, leading to predictable mixing time. A symmetric Metropolis-Klein (SMK) algorithm is also proposed, which is proven to be geometrically ergodic.
\end{abstract}

\IEEEpeerreviewmaketitle

\vspace{0.5em}

\IEEEpeerreviewmaketitle

\section{Introduction}
\newtheorem{my1}{Lemma}
\newtheorem{my2}{Theorem}
\newtheorem{my3}{Definition}
\newtheorem{my4}{Proposition}
\newtheorem{my5}{Corollary}

Recently, the lattice Gaussian distribution has emerged as a common theme in
various research domains. In mathematics, Banaszczyk firstly applied it to prove the transference theorems for lattices \cite{Banaszczyk}. In coding,
lattice Gaussian distribution was employed to obtain the full shaping gain for lattice coding \cite{Forney_89,Kschischang_Pasupathy}, and to achieve the capacity of the Gaussian channel \cite{LB_13}. It was also used to achieve information-theoretic security in the Gaussian wiretap channel \cite{LLBS_12,7360779} and in the bidirectional relay channel \cite{7058433}, respectively. In cryptography, the lattice Gaussian distribution has already become a central tool in the construction of many primitives. Specifically, Micciancio and Regev used it to propose lattice-based cryptosystems based on the worst-case hardness assumptions \cite{MicciancioGaussian}. Meanwhile, it also has underpinned the fully-homomorphic encryption for cloud computing \cite{GentryDissertation}. Algorithmically, lattice Gaussian sampling with a suitable variance allows to solve the shortest vector problem (SVP) and the closest vector problem (CVP) \cite{RegevSolvingtheShortestVectorProblem,RegevSolvingtheClosestVectorProblem}; for example, it has led to efficient lattice decoding for multi-input multi-output (MIMO) systems \cite{CongRandom,DerandomizedJ}. In theory, it has been demonstrated that lattice Gaussian sampling is equivalent to CVP via a polynomial-time dimension-preserving reduction \cite{DGStoCVPSVP}, and SVP is essentially a special case of the CVP.

Due to the central role of the lattice Gaussian distribution playing in these fields, its sampling algorithms become an important computational problem. In contrast to sampling from a continuous Gaussian distribution, it is by no means trivial to perform the sampling even from a low-dimensional discrete Gaussian distribution. As the default sampling algorithm for lattices, Klein's algorithm \cite{Klein} is capable to sample from the lattice Gaussian distribution within a negligible statistical distance if and only if the standard deviation $\sigma\geq \sqrt{\omega(\text{log}\ n)}\cdot\text{max}_{1\leq i \leq n}\|\mathbf{\widehat{b}}_i\|$ \cite{Trapdoor}, where $\omega(\text{log}\ n)$ is a superlogarithmic function, $n$ denotes the lattice dimension and $\mathbf{\widehat{b}}_i$'s are the Gram-Schmidt vectors of the lattice basis $\mathbf{B}$. However, such requirement of $\sigma$ can be excessively large, rendering Klein's algorithm inapplicable to many scenarios of interest.

Markov chain Monte Carlo (MCMC) methods attempt to sample from the target distribution by building a Markov chain, which randomly generates the next sample conditioned on previous samples.
After a burn-in period, which is normally measured by the \emph{mixing time}, the Markov chain will reach a stationary distribution, and successful sampling from the complex target distribution can be carried out. To this end, the Gibbs algorithm was introduced into lattice Gaussian sampling, which employs univariate conditional sampling to build a Markov chain \cite{ZhengWangMCMCLatticeGaussian}. It is able to sample beyond the range of Klein's algorithm. In \cite{ZhengWangMCMCLatticeGaussian}, a flexible block-based Gibbs algorithm was also presented, which performs sampling over multiple elements within a block. In this way, the correlation within the block could be exploited, leading to a faster convergence especially in the case of highly correlated components. Unfortunately, related analysis of the convergence rate for the associated Markov chains in these two algorithms was lacking, resulting in an unpredictable mixing time.

On the other hand, Gibbs sampling has already been adapted to signal detection for multi-input multi-output (MIMO) communications \cite{HassibiMCMCnew,McmcDatta,MCMCBehrouz,ChenMCMC,XiaodongWangMultilevel,MCMCHaidongZhu}. In particular, the selection of $\sigma$ (also referred to as ``temperature'') is studied in \cite{HassibiMCMCnew} and it is argued that $\sigma$ should grow as fast as the signal-to-noise ratio (SNR) in general. In \cite{McmcDatta}, a mixed-Gibbs sampler is proposed to achieve near-optimal performance, which takes the advantages of an efficient stopping criterion and a multiple restart strategy. Moreover, Gibbs sampling is also introduced into soft-output decoding in MIMO systems, where the extrinsic information calculated by a priori probability (APP) detector is used to produce soft outputs \cite{MCMCBehrouz}. In \cite{ChenMCMC}, an investigation of Gibbs-based MCMC receivers
in different communication channels are given. Due to the finite state space formed by a finite modulation constellation, those Gibbs samplers converge exponentially fast to the stationary distribution. However, the rate of convergence has not yet been determined.

In this paper, another famous MCMC scheme, known as the Metropolis-Hastings (MH) algorithm \cite{Hastings1970}, is studied in detail for lattice Gaussian sampling. In particular, it makes use of a \emph{proposal distribution} which suggests a possible state candidate and then employs an acceptance-rejection rule to decide whether to accept the suggested candidate in the next Markov move. Obviously, the art of designing an efficient MH algorithm lies in choosing an appropriate proposal distribution, and this motivates us to design the target proposal distributions based on Klein's algorithm.

In the proposed independent Metropolis-Hastings-Klein (MHK) algorithm, a candidate at each Markov move is generated from a Gaussian-like proposal distribution via Klein's algorithm. In this case, we show that the Markov chain induced by the independent MHK algorithm is uniformly ergodic, which implies it converges exponentially fast to the stationary distribution irrespective of the starting state. Its convergence rate is then estimated given the lattice basis $\mathbf{B}$, the query point $\mathbf{c}$ and the standard derivation $\sigma$. Thus, the mixing time of the induced Markov chain becomes predictable. To the best of our knowledge, this is the first time that the convergence rate of MCMC in communications and signal processing is determined analytically since MCMC was introduced into this field in 1990's \cite{WangxiaodongMCMag}.

Different from the algorithms in \cite{RegevSolvingtheShortestVectorProblem,RegevSolvingtheClosestVectorProblem} which have exponential space and time complexity, the proposed independent MHK algorithm has polynomial space complexity, and its time complexity\footnote{In this paper, the computational complexity is measured by the number of arithmetic operations (additions, multiplications, comparisons, etc.). The time complexity of an MCMC sampler can be estimated by the mixing time times the complexity of each Markov move.} varies with $\sigma$, where a larger value of $\sigma$ corresponds to smaller mixing time. This is in agreement with the fact we knew before: if $\sigma$ is large enough, then there is no need of MCMC in lattice Gaussian sampling since Klein's algorithm can be applied directly with polynomial time complexity.

The second proposed algorithm, namely the symmetric Metropolis-Klein (SMH) algorithm, establishes a symmetric proposal distribution between two consecutive Markov states. We show it also converges to the stationary distribution exponentially fast but the selection of the initial state also plays a role. Such a case is referred to as geometric ergodicity in MCMC literature \cite{RobertsGeneralstatespace}. Besides the geometric ergodicity, another advantage of the proposed SMH algorithm lies in its remarkable elegance and simplicity, which comes from the usage of a symmetrical proposal distribution.
\begin{figure}[t]
\vspace{-2em}
\begin{center}
\hspace{-1em}\includegraphics[width=4in,height=2.4in]{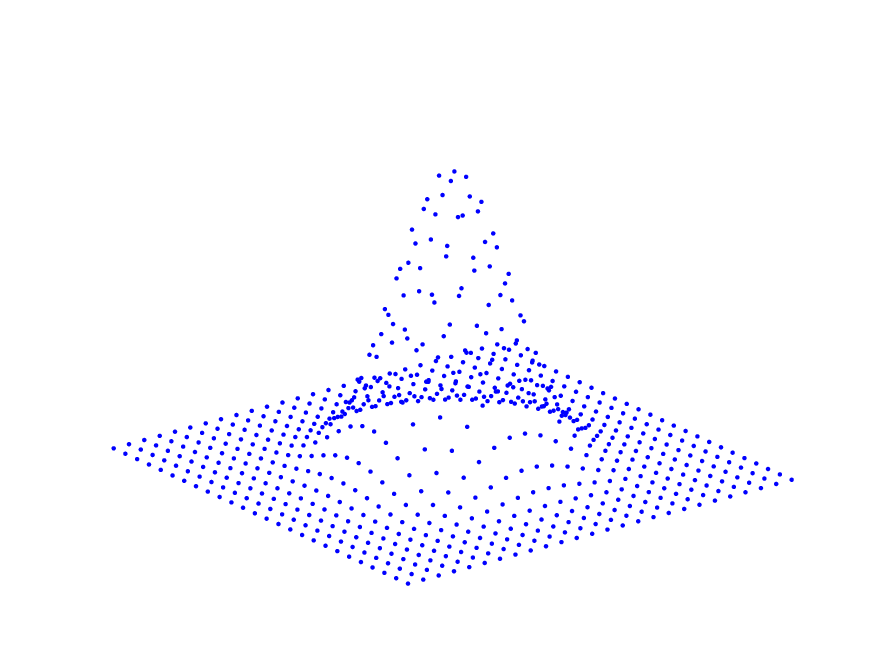}
\end{center}
\vspace{-2em}
  \caption{Illustration of a two-dimensional lattice Gaussian distribution.}
  \label{simulation x}
\end{figure}

To summarize, the main contributions of this paper are the following:
\begin{description}
  \item[1)] The independent MHK algorithm is proposed for lattice Gaussian sampling, where the Markov chain arising from it converges exponentially fast to the stationary distribution.
  \item[2)] The convergence rate of the independent MHK algorithm is derived explicitly in terms of the theta series, thereby making the estimation of mixing time possible.
  \item[3)] The SMH algorithm is further proposed for lattice Gaussian sampling, which not only achieves exponential convergence, but also is simpler due to its symmetry.
\end{description}

The rest of this paper is organized as follows. Section II introduces the lattice Gaussian distribution
and briefly reviews the basics of MCMC methods. In Section III, we propose the independent MHK algorithm for lattice Gaussians, where uniform ergodicity is demonstrated. In Section IV, the convergence rate of the independent MHK algorithm is analyzed and explicitly calculated in terms of the theta series. In Section V, the proposed SMH algorithm for lattice Gaussian sampling is given, followed by the demonstration of geometric ergodicity. Finally, Section VI concludes the paper.

\emph{Notation:} Matrices and column vectors are denoted by upper
and lowercase boldface letters, and the transpose, inverse, pseudoinverse
of a matrix $\mathbf{B}$ by $\mathbf{B}^T, \mathbf{B}^{-1},$ and
$\mathbf{B}^{\dag}$, respectively. We denote by $\mathbf{b}_i$ the $i$th
column of the matrix $\mathbf{B}$, by $\mathbf{\widehat{b}}_i$ the $i$th
Gram-Schmidt vector of $\mathbf{B}$, and by $b_{i,j}$ the entry in the $i$th row
and $j$th column of $\mathbf{B}$. $\lceil x \rfloor$ denotes rounding to
the integer closest to $x$. If $x$ is a complex number, $\lceil x \rfloor$
rounds the real and imaginary parts separately. In addition, we use the standard \emph{small omega} notation $\omega(\cdot)$, i.e., $|\omega(g(n))|>k\cdot|g(n)|$ for every fixed positive number $k>0$.

\section{Preliminaries}
In this section, we introduce the background and mathematical tools needed
to describe and analyze the proposed lattice Gaussian sampling algorithms.

\subsection{Lattice Gaussian Distribution}
Let $\mathbf{B}=[\mathbf{b}_1,\ldots,\mathbf{b}_n]\subset \mathbb{R}^n$ consist of $n$ linearly independent vectors. The $n$-dimensional lattice $\Lambda$ generated by $\mathbf{B}$ is defined by
\begin{equation}
\Lambda=\{\mathbf{Bx}: \mathbf{x}\in \mathbb{Z}^n\},
\end{equation}
where $\mathbf{B}$ is called the lattice basis. We define the Gaussian function centered at $\mathbf{c}\in \mathbb{R}^n$ for standard deviation $\sigma>0$ as
\begin{equation}
\rho_{\sigma, \mathbf{c}}(\mathbf{z})=e^{-\frac{\|\mathbf{z}-\mathbf{c}\|^2}{2\sigma^2}},
\end{equation}
for all $\mathbf{z}\in\mathbb{R}^n$. When $\mathbf{c}$ or $\sigma$ are not specified, we assume that they are $\mathbf{0}$ and $1$ respectively. Then, the \emph{discrete Gaussian distribution} over $\Lambda$ is defined as
\begin{equation}
D_{\Lambda,\sigma,\mathbf{c}}(\mathbf{x})=\frac{\rho_{\sigma, \mathbf{c}}(\mathbf{Bx})}{\rho_{\sigma, \mathbf{c}}(\Lambda)}=\frac{e^{-\frac{1}{2\sigma^2}\parallel \mathbf{Bx}-\mathbf{c} \parallel^2}}{\sum_{\mathbf{x} \in \mathbb{Z}^n}e^{-\frac{1}{2\sigma^2}\parallel \mathbf{Bx}-\mathbf{c} \parallel^2}}
\label{lattice gaussian distribution}
\end{equation}
for all $\mathbf{x}\in \mathbb{Z}^n$, where $\rho_{\sigma, \mathbf{c}}(\Lambda)\triangleq \sum_{\mathbf{\mathbf{Bx}}\in\Lambda}\rho_{\sigma, \mathbf{c}}(\mathbf{Bx})$ is just a scaling to obtain a probability distribution.

Note that this definition differs slightly from the one in \cite{MicciancioGaussian}, where $\sigma$ is
scaled by a constant factor $\sqrt{2\pi}$ (i.e., $s =\sqrt{2\pi}\sigma$). Fig. 1 illustrates the discrete Gaussian distribution over $\mathbb{Z}^2$. As can be seen clearly, it resembles a continuous Gaussian distribution, but is only defined over a lattice. In fact, discrete and continuous
Gaussian distributions share similar properties, if the \emph{flatness
factor} is small \cite{LLBS_12}.

\renewcommand{\algorithmicrequire}{\textbf{Input:}}  
\renewcommand{\algorithmicensure}{\textbf{Output:}} 

\begin{algorithm}[t]
\caption{Klein's Algorithm}
\begin{algorithmic}[1]
\Require
$\mathbf{B}, \sigma, \mathbf{c}$
\Ensure
$\mathbf{Bx}\in\Lambda$
\State let $\mathbf{B}=\mathbf{QR}$ and $\mathbf{c'}=\mathbf{Q}^{\dag}\mathbf{c}$
\For {$i=n$,\ \ldots,\ 1}
\State let $\sigma_i=\frac{\sigma}{|r_{i,i}|}$ and $\widetilde{x}_i=\frac{c'_i-\sum^n_{j=i+1}r_{i,j}x_j}{r_{i,i}}$
\State sample $x_i$ from $D_{\mathbb{Z},\sigma_i,\widetilde{x}_i}$
\EndFor
\State return $\mathbf{Bx}$
\end{algorithmic}
\end{algorithm}

\subsection{Klein's Algorithm}
Intuitively, the shape of $D_{\Lambda,\sigma,\mathbf{c}}(\mathbf{x})$ suggests that a lattice point $\mathbf{Bx}$ closer to $\mathbf{c}$ will be sampled with a higher probability. Therefore, sampling from the lattice Gaussian distribution can be naturally used to solve the CVP (where $\mathbf{c}$ is the query point) and SVP (where $\mathbf{c}=\mathbf{0}$) in lattices. Because of this, Klein's algorithm that samples from a Gaussian-like distribution was originally proposed for lattice decoding \cite{Klein}.

As shown in Algorithm 1, the operation of the Klein's algorithm has polynomial complexity $O(n^2)$ excluding QR decomposition (which may be done only once in the beginning). More precisely, by sequentially sampling from the 1-dimensional conditional Gaussian distribution $D_{\mathbb{Z},\sigma_i,\widetilde{x}_i}$ in a backward order from $x_n$ to $x_1$, the Gaussian-like distribution arising from Klein's algorithm is given by
\begin{eqnarray}
P_{\text{Klein}}(\mathbf{x})&=&\prod^n_{i=1}D_{\mathbb{Z},\sigma_i,\widetilde{x}_i}(x_i)=\frac{\rho_{\sigma, \mathbf{c}}(\mathbf{Bx})}{\prod^n_{i=1}\rho_{\sigma_i, \widetilde{x}_i}(\mathbb{Z})}\notag\\
&=&\frac{e^{-\frac{1}{2\sigma^2}\parallel \mathbf{Bx}-\mathbf{c} \parallel^2}}{\prod^n_{i=1}\sum_{\widetilde{x}_i\in\mathbb{Z}}e^{-\frac{1}{2\sigma^2_i}\|x_i-\widetilde{x}_i\|^2}},
\label{klein distribution}
\end{eqnarray}
where $\widetilde{x}_i=\frac{c'_i-\sum^n_{j=i+1}r_{i,j}x_j}{r_{i,i}}$, $\sigma_i=\frac{\sigma}{|r_{i,i}|}=\frac{\sigma}{\|\mathbf{\widehat{b}}_i\|}$, $\mathbf{c'}=\mathbf{Q}^{\dag}\mathbf{c}$, $r_{i,j}$ denotes the element of the upper triangular matrix $\mathbf{R}$ from the QR decomposition $\mathbf{B}=\mathbf{QR}$ and $\mathbf{\widehat{b}}_i$'s are the Gram-Schmidt vectors of $\mathbf{B}$ with $\|\mathbf{\widehat{b}}_i\|=|r_{i,i}|$.

Furthermore, it has been demonstrated in \cite{Trapdoor} that $P_{\text{Klein}}(\mathbf{x})$ is close to $D_{\Lambda,\sigma,\mathbf{c}}(\mathbf{x})$ within a negligible statistical distance if
\begin{equation}
\sigma\geq\sqrt{\omega(\text{log}\ n)}\cdot\max_{1\leq i \leq n}\|\mathbf{\widehat{b}}_i\|,
\end{equation}
However, even with the help of lattice reduction\footnote{It is well known lattice reduction such as the LLL algorithm is able to significantly improve $\min_i\|\widehat{\mathbf{b}}_i\|$ while reducing $\max_i\|\widehat{\mathbf{b}}_i\|$ at the same time \cite{LLLoriginal}.} (e.g., LLL reduction), the variance $\sqrt{\omega(\text{log}\ n)}\cdot\max_{1\leq i \leq n}\|\mathbf{\widehat{b}}_i\|$ can be too large to be useful.

\subsection{MCMC Methods}
As for the lattice Gaussian sampling in the range $\sigma<\sqrt{\omega(\text{log}\ n)}\cdot\max_{1\leq i \leq n}\|\mathbf{\widehat{b}}_i\|$, MCMC methods have become an alternative solution, where the discrete Gaussian distribution $D_{\Lambda,\sigma,\mathbf{c}}$ is viewed as a complex target distribution lacking direct sampling methods. By establishing a Markov chain that randomly generates the next state, MCMC is capable of sampling from the target distribution of interest, thereby removing the restriction on $\sigma$ \cite{ZhengWangMCMCLatticeGaussian}.

As an important parameter which measures the time required by a
Markov chain to get close to its stationary distribution, the \emph{mixing time} is defined as \cite{mixingtimemarkovchain}
\begin{equation}
t_{\text{mix}}(\epsilon)=\text{min}\{t:\max\|P^t(\mathbf{x}, \cdot)-\pi(\cdot)\|_{TV}\leq \epsilon\},
\label{mixing time}
\end{equation}
where $\|\cdot\|_{TV}$ represents the total variation distance (other measures of distance also exist, see \cite{RosenthalOPDAMCMC} for more details). It is well known that the \emph{spectral gap} $\gamma$ of the transition matrix offers an upper bound on the mixing time
\begin{equation}
t_{\text{mix}}(\epsilon)\leq \frac{1}{\gamma}\text{log}\left(\frac{1}{\pi_{\text{min}}\epsilon}\right),
\label{mixing times}
\end{equation}
where $\pi_{\text{min}}=\text{min}_{\mathbf{x\in\Omega}}\pi(\mathbf{x})$, $\Omega$ stands for the state space, $\gamma=1-|\lambda_1|>0$ and $\lambda_1$ represents the second largest eigenvalue of the transition matrix $\mathbf{P}$ in a Markov chain. Therefore, a large value of the spectral gap leads to rapid convergence to stationarity \cite{RapidlyMixingRandall}.

However, the spectrum of a Markov chain can be hard to analyze, especially when the state space $\Omega$ becomes exponentially large, making it difficult to have a compact mathematical expression of the adjacency matrix. Thanks to the celebrated \emph{coupling technique}, for any Markov chain with finite state space $\Omega$, exponentially fast convergence can be demonstrated if the underlying Markov chain is irreducible and aperiodic with an invariant distribution $\pi$ \cite{mixingtimemarkovchain}. Nevertheless, in the case of lattice Gaussian sampling, the countably infinite state space $\mathbf{x}\in \mathbb{Z}^n$ naturally becomes a challenge. For this reason, we perform the convergence analysis from the beginning --- ergodicity \cite{Meynbook}.

\begin{my3}
Let $\mathbf{P}$ be an irreducible and aperiodic transition matrix for a Markov chain. If the chain is positive recurrent, then it is ergodic, namely, there is a unique probability distribution $\pi$ on $\Omega$ and for all $\mathbf{x}\in\Omega$,
\begin{equation}
\underset{t\rightarrow\infty}{{\lim}}\|P^t(\mathbf{x},\cdot)-\pi\|_{TV}=0,
\end{equation}
{where $P^t(\mathbf{x}; \cdot)$ denotes a row of the transition matrix $\mathbf{P}$ for $t$ Markov moves.}
\end{my3}

Although \emph{ergodicity} implies asymptotic convergence to stationarity, it does not say anything about the convergence rate. To this end, the following definition is given \cite{Meynbook}.

\begin{my3}
A Markov chain with stationary distribution $\pi(\cdot)$ is uniformly ergodic if there exists $0<\delta<1$ and $M<\infty$ such that for all $\mathbf{x}$
\begin{equation}
\|P^t(\mathbf{x}, \cdot)-\pi(\cdot)\|_{TV}\leq M(1-\delta)^t.
\end{equation}
\end{my3}

Obviously, the exponential decay coefficient $\delta$ is key to determine the convergence rate. As $M$ is a constant, the convergence rate does not depend on the initial state $\mathbf{x}$. As a weaker version of ergodicity, \emph{geometric ergodicity} also converges exponentially, but $M$ is parameterized by the initial state $\mathbf{x}$.

\begin{my3}
A Markov chain with stationary distribution $\pi(\cdot)$ is geometrically ergodic if there exists $0<\delta<1$ and $M(\mathbf{x})<\infty$ such that for all $\mathbf{x}$
\begin{equation}
\|P^t(\mathbf{x}, \cdot)-\pi(\cdot)\|_{TV}\leq M(\mathbf{x})(1-\delta)^t.
\label{geo-ergodic}
\end{equation}
\end{my3}

Besides exponential convergence, polynomial convergence also exists \cite{JarnerPolynomialconvergence}, which goes beyond the scope of this paper due to its slow convergence. Unless stated otherwise, the state space of the Markov chain we are concerned with throughout the context is the countably infinite $\Omega=\mathbb{Z}^n$.

\subsection{Classical MH Algorithms}
The origin of the Metropolis algorithm can be traced back to the celebrated work of \cite{MetropolisOrignial} in 1950's. In \cite{Hastings1970}, the original Metropolis algorithm was successfully extended to a more general scheme known as the Metropolis-Hastings (MH) algorithm. In particular, let us consider a target invariant distribution $\pi$ together with a proposal distribution
$q(\mathbf{x},\mathbf{y})$. Given the current state $\mathbf{x}$ for Markov chain $\mathbf{X}_t$, a state candidate $\mathbf{y}$ for the next Markov move $\mathbf{X}_{t+1}$ is generated from the proposal distribution $q(\mathbf{x},\mathbf{y})$. Then the acceptance ratio $\alpha$ is computed by
\begin{equation}
\alpha(\mathbf{x},\mathbf{y})=\text{min}\left\{1,\frac{\pi(\mathbf{y})q(\mathbf{y},\mathbf{x})}{\pi(\mathbf{x})q(\mathbf{x},\mathbf{y})}\right\},
\label{quantity compute}
\end{equation}
and $\mathbf{y}$ will be accepted as the new state by $\mathbf{X}_{t+1}$ with
probability $\alpha$. Otherwise, $\mathbf{x}$ will be retained by $\mathbf{X}_{t+1}$. In this way, a Markov chain $\{\mathbf{X}_0, \mathbf{X}_1, \ldots\}$ is established with the transition probability $P(\mathbf{x},\mathbf{y})$ as follows:
\begin{equation}
P(\mathbf{x},\mathbf{y})=\begin{cases}q(\mathbf{x},\mathbf{y})\alpha(\mathbf{x},\mathbf{y}) \ \ \ \ \ \ \ \ \ \ \ \ \ \ \text{if}\ \mathbf{y}\neq\mathbf{x}, \\
       1-\sum_{\mathbf{z}\neq\mathbf{x}}q(\mathbf{x},\mathbf{z})\alpha(\mathbf{x},\mathbf{z})\ \ \text{if}\ \mathbf{y}=\mathbf{x}.
       \end{cases}
\label{eqn:RandomDiscrete}
\end{equation}

It is interesting that in MH algorithms, the proposal distribution $q(\mathbf{x},\mathbf{y})$ can be any fixed distribution from which we can conveniently draw samples. Undoubtedly, the fastest converging proposal distribution would be $q(\mathbf{x},\mathbf{y})=\pi(\mathbf{y})$ itself, but in most cases of interest $\pi$ cannot be sampled directly. To this end, many variations of MH algorithms with different configurations of $q(\mathbf{x},\mathbf{y})$ were proposed.

\section{Independent MHK Algorithm}
In this section, the independent Metropolis-Hastings-Klein (MHK) algorithm for lattice Gaussian sampling is firstly presented. Then, we show that the Markov chain induced by the proposed algorithm is uniformly ergodic.

\subsection{Independent MHK Algorithm}
In the proposed independent MHK algorithm, Klein's sampling is used to generate the state candidate $\mathbf{y}$ for the each Markov move $\mathbf{X}_{t+1}$. As shown in Algorithm 2, it consists of three basic steps:

1)\ \ \hspace{-.2em}\emph{Sample from the independent proposal distribution with Klein's algorithm to obtain the candidate state $\mathbf{y}$ for $\mathbf{X}_{t+1}$},
\begin{eqnarray}
q(\mathbf{x},\mathbf{y})&=&q(\mathbf{y})\ =\ P_{\text{Klein}}(\mathbf{y})\notag\\
&=&\frac{\rho_{\sigma, \mathbf{c}}(\mathbf{By})}{\prod^n_{i=1}\rho_{\sigma_i, \widetilde{y}_i}(\mathbb{Z})}\notag\\
&=&\frac{e^{-\frac{1}{2\sigma^2}\parallel \mathbf{By}-\mathbf{c} \parallel^2}}{\prod^n_{i=1}\sum_{\widetilde{y}_i\in\mathbb{Z}}e^{-\frac{1}{2\sigma^2_i}\|y_i-\widetilde{y}_i\|^2}}
\label{MH proposal density}
\end{eqnarray}
where $\mathbf{y}\in \mathbb{Z}^n$, $\widetilde{y}_i=\frac{c'_i-\sum^n_{j=i+1}r_{i,j}y_j}{r_{i,i}}$, $\sigma_i=\frac{\sigma}{|r_{i,i}|}=\frac{\sigma}{\|\mathbf{\widehat{b}}_i\|}$, $\mathbf{c'}=\mathbf{Q}^{\dag}\mathbf{c}$, $\mathbf{B}=\mathbf{QR}$ by QR decomposition and $\mathbf{\widehat{b}}_i$'s are the Gram-Schmidt vectors of $\mathbf{B}$.

2)\ \ \hspace{-.2em}\emph{Calculate the acceptance ratio $\alpha(\mathbf{x},\mathbf{y})$}
\begin{eqnarray}
\alpha(\mathbf{x},\mathbf{y})\hspace{-.1em}&=&\hspace{-.1em}\text{min}\hspace{-.2em}\left\{\hspace{-.1em}1,\hspace{-.1em}\frac{\pi(\mathbf{y})q(\mathbf{y},\mathbf{x})}{\pi(\mathbf{x})q(\mathbf{x},\mathbf{y})}\hspace{-.1em}\right\}\hspace{-.1em}=\hspace{-.1em}\text{min}\hspace{-.2em}\left\{\hspace{-.1em}1,\hspace{-.1em}\frac{\pi(\mathbf{y})q(\mathbf{x})}{\pi(\mathbf{x})q(\mathbf{y})}\hspace{-.1em}\right\}\notag\\
&=&\text{min}\hspace{-.2em}\left\{\hspace{-.1em}1,\hspace{-.1em}\frac{\prod^n_{i=1}\rho_{\sigma_i, \widetilde{y}_i}(\mathbb{Z})}{\prod^n_{i=1}\rho_{\sigma_i, \widetilde{x}_i}(\mathbb{Z})}\hspace{-.1em}\right\},
\label{MH quantity compute}
\end{eqnarray}
where $\pi=D_{\Lambda,\sigma,\mathbf{c}}$.

3)\ \ \hspace{-.2em}\emph{Make a decision for $\mathbf{X}_{t+1}$ based on $\alpha(\mathbf{x},\mathbf{y})$ to accept $\mathbf{X}_{t+1}=\mathbf{y}$ or not.}

A salient feature of the independent MHK algorithm is that the generation of the state candidate $\mathbf{y}$ is independent of the previous one, which is completely accomplished by Klein's algorithm. Therefore, the connection between two consecutive Markov states only lies in the decision part. The complexity of the MCMC sampler is given by the number of Markov moves times the complexity of each move, i.e., $O(t_{\mathrm{mix}}\cdot n^2)$.

It is easy to verify that the Markov chain with the independent proposal distribution $q$ shown in (\ref{MH proposal density}) is irreducible, aperiodic and positive recurrent, which naturally leads to an ergodic Markov chain \cite{mixingtimemarkovchain}. Then, we have the following well-known result, whose proof can be found in \cite{mixingtimemarkovchain}.

\begin{my4}
Given the target lattice Gaussian distribution $\pi=D_{\Lambda,\sigma,\mathbf{c}}$, the Markov chain induced by the independent MHK algorithm is ergodic:
\begin{equation}
\underset{t\rightarrow\infty}{{\lim}}\|P^t(\mathbf{x}; \cdot)-D_{\Lambda,\sigma,\mathbf{c}}(\cdot)\|_{TV}=0
\end{equation}
{for all states $\mathbf{x}\in\mathbb{Z}^n$.}
\end{my4}

\subsection{Uniform Ergodicity}
The independent proposal distribution defined in (\ref{MH proposal density}) enjoys the following property.
\begin{my1}
In the independent MHK algorithm for lattice Gaussian sampling from $D_{\Lambda,\sigma,\mathbf{c}}$, there exists $\delta>0$ such that
\begin{equation}
\frac{q(\mathbf{x})}{\pi(\mathbf{x})}\geq \delta
\label{xxxx}
\end{equation}
for all $\mathbf{x}\in \mathbb{Z}^n$, where $q(\mathbf{x})=P_{\mathrm{Klein}}(\mathbf{x})$.
\end{my1}
\begin{proof}
Using (\ref{lattice gaussian distribution}) and (\ref{klein distribution}), we have
\begin{eqnarray}
\frac{q(\mathbf{x})}{\pi(\mathbf{x})}&=&\frac{\rho_{\sigma, \mathbf{c}}(\mathbf{Bx})}{\prod^n_{i=1}\rho_{\sigma_i, \widetilde{x}_i}(\mathbb{Z})}\cdot\frac{\rho_{\sigma, \mathbf{c}}(\mathbf{\Lambda})}{\rho_{\sigma, \mathbf{c}}(\mathbf{Bx})}\notag\\
&=&\frac{\rho_{\sigma, \mathbf{c}}(\mathbf{\Lambda})}{\prod^n_{i=1}\rho_{\sigma_i, \widetilde{x}_i}(\mathbb{Z})}\label{x2}\notag\\
&\overset{(a)}{\geq}&\frac{\rho_{\sigma, \mathbf{c}}(\mathbf{\Lambda})}{\prod^n_{i=1}\rho_{\sigma_i}(\mathbb{Z})}=\delta\label{x1}
\end{eqnarray}
where $(a)$ holds due to the fact that \cite{MicciancioGaussian}
\begin{equation}
\rho_{\sigma_i, \tilde{x}}(\mathbb{Z})\leq \rho_{\sigma_i}(\mathbb{Z}) \triangleq \sum_{j\in \mathbb{Z}}e^{-\frac{1}{2\sigma_i^2} j^2}.
\label{nnnnnn}
\end{equation}

As can be seen clearly, the right-hand side (RHS) of (\ref{x1}) is completely independent
of $\mathbf{x}$, meaning it can be expressed as a constant $\delta$ determined by the given
$\mathbf{B}$, $\mathbf{c}$ and $\sigma$. Therefore, the proof is completed.
\end{proof}

We then arrive at a main Theorem to show the uniform ergodicity of the proposed algorithm.

\begin{my2}
Given the invariant lattice Gaussian distribution $D_{\Lambda,\sigma,\mathbf{c}}$, the Markov chain established by the independent MHK algorithm is uniformly ergodic:
\begin{equation}
\|P^t(\mathbf{x}, \cdot)-D_{\Lambda,\sigma,\mathbf{c}}(\cdot)\|_{TV}\leq (1-\delta)^t
\end{equation}
for all $\mathbf{x}\in \mathbb{Z}^n$.
\end{my2}

\begin{proof}
By (\ref{MH proposal density}) and (\ref{MH quantity compute}), the transition probability $P(\mathbf{x},\mathbf{y})$ of the independent MHK algorithm is given by
\begin{equation}\label{eq:20}
\hspace{-.3em}P(\mathbf{x},\hspace{-.1em}\mathbf{y})\hspace{-.2em}=\hspace{-.4em}\begin{cases}\hspace{-.2em}\text{min}\left\{q(\mathbf{y}),\frac{\pi(\mathbf{y})q(\mathbf{x})}{\pi(\mathbf{x})}\right\} \ \ \ \ \ \ \ \ \ \ \ \ \ \hspace{.6em}\ \text{if}\ \hspace{-.1em}\mathbf{y}\hspace{-.2em}\neq\hspace{-.1em}\mathbf{x}, \\
\hspace{-.2em}q(\mathbf{x})\hspace{-.2em}+\hspace{-.4em}\underset{\mathbf{z}\neq\mathbf{x}}{\sum}\hspace{-.1em}\max\hspace{-.1em}\left\{\hspace{-.1em}0, \hspace{-.1em}q(\mathbf{z})\hspace{-.1em}-\hspace{-.1em}\frac{\pi(\mathbf{z})q(\mathbf{x})}{\pi(\mathbf{x})}\hspace{-.2em}\right\}       \hspace{-.1em}\ \text{if}\ \hspace{-.1em}\mathbf{y}\hspace{-.2em}=\hspace{-.1em}\mathbf{x}.
       \end{cases}
\end{equation}

Using Lemma 1, it is straightforward to check that the following relationship holds
\begin{equation}
P(\mathbf{x}, \mathbf{y})\geq\delta\pi(\mathbf{y})
\label{mmmmm}
\end{equation}
for all $\mathbf{x},\mathbf{y}\in \mathbb{Z}^n$. Now, consider a coupling of two Markov chains $\mathbf{X}_{t}$ and $\mathbf{X}'_{t}$, which marginally update according to the same transition probability \eqref{eq:20}.
$\mathbf{X}'_{t}$ is supposed to start from the stationary distribution $\pi$, and $\mathbf{X}_{t}$ from an initial state $\mathbf{x}_0$, which is not necessarily stationary.

According to the \emph{coupling inequality} \cite{mixingtimemarkovchain}, the variation distance between the distributions of $\mathbf{X}_{t}$ and $\mathbf{X}'_{t}$ is upper bounded by
\begin{equation}
\|P^t(\mathbf{x}_0,\cdot)-\pi(\cdot)\|_{TV}\leq P(\mathbf{X}_t\neq \mathbf{X}'_t).
\label{aaaaa}
\end{equation}
On the other hand, any coupling of Markov chains can be modified so that the two chains stay together at all times once they meet at a same state \cite{mixingtimemarkovchain}, namely,
\begin{equation}
\text{if}\ \mathbf{X}_{n}=\mathbf{X}'_{n},\ \text{then}\ \mathbf{X}_{t}=\mathbf{X}'_{t}\ \text{for}\ t\geq n.
\end{equation}
Therefore, given the event $\mathbf{X}_t\neq\mathbf{X}'_t$, there is no coupling in any of the $t$ consecutive moves. By \eqref{mmmmm}, for each move we have probability at least $\delta$ of
making $\mathbf{X}_i$ and $\mathbf{X}'_i$ ($i=1,2,\ldots,t$) equal and we have
\begin{eqnarray}
P(\mathbf{X}_t\neq\mathbf{X}'_t)\hspace{-1em}&=&\hspace{-1em}P(\mathbf{X}_t\neq\mathbf{X}'_t, \ldots, \mathbf{X}_{0}\neq\mathbf{X}'_{0})\notag\\
\hspace{-1em}&=&\hspace{-1em}\prod_{i=1}^t\hspace{-.2em}P(\mathbf{X}_i\neq\mathbf{X}'_i|\mathbf{X}_{i-1}\neq\mathbf{X}'_{i-1}\hspace{-.1em}) \hspace{-.1em}\cdot\hspace{-.1em} P(\mathbf{X}_0\neq\mathbf{X}'_0)\notag\\
\hspace{-1em}&\leq&\hspace{-1em}\prod_{i=1}^tP(\mathbf{X}_i\neq\mathbf{X}'_i|\mathbf{X}_{i-1}\neq\mathbf{X}'_{i-1})\notag\\
\hspace{-1em}&=&\hspace{-1em}\prod_{i=1}^t\left[1-P(\mathbf{X}_i=\mathbf{X}'_i|\mathbf{X}_{i-1}\neq\mathbf{X}'_{i-1})\right]\notag\\
\hspace{-1em}&=&\hspace{-1em}\left[\hspace{-.1em}1\hspace{-.2em}-\hspace{-.2em}\sum_{\mathbf{y}\in\mathbb{Z}^n} \hspace{-.3em}P(\mathbf{X}_i=\mathbf{X}'_i=\mathbf{y}|\mathbf{X}_{i-1}\neq\mathbf{X}'_{i-1})\right]^t\notag\\
\hspace{-1em}&\overset{(b)}{\leq}&\hspace{-1em}\left[1-\sum_{\mathbf{y}\in\mathbb{Z}^n}\delta\pi(\mathbf{y})\right]^t\notag\\
\hspace{-1em}&=&\hspace{-1em}(1-\delta)^t,
\label{pppppppp}
\end{eqnarray}
where $(b)$ is due to \eqref{mmmmm}.

\begin{algorithm}[t]
\caption{Independent Metropolis-Hastings-Klein Algorithm for Lattice Gaussian Sampling}
\begin{algorithmic}[1]
\Require
$\mathbf{B}, \sigma, \mathbf{c}, \mathbf{X}_0, t_{\text{mix}}(\epsilon)$
\Ensure sample from a distribution statistically close to $\pi=D_{\Lambda,\sigma,\mathbf{c}}$
\For {$t=$1,2,\ \ldots,\ }
\State let $\mathbf{x}$ denote the state of $\mathbf{X}_{t-1}$
\State generate $\mathbf{y}$ from the proposal distribution $q(\mathbf{x},\mathbf{y})$ in (\ref{MH proposal density})
\State calculate the acceptance ratio $\alpha(\mathbf{x},\mathbf{y})$ in (\ref{MH quantity compute})
\State generate a sample $u$ from the uniform density $U[0,1]$
\If {$u\leq \alpha(\mathbf{x},\mathbf{y})$}
\State let $\mathbf{X}_t=\mathbf{y}$
\Else
\State $\mathbf{X}_t=\mathbf{x}$
\EndIf
\If {$t\geq t_{\text{mix}}(\epsilon)$ }
\State output the state of $\mathbf{X}_t$
\EndIf
\EndFor
\end{algorithmic}
\end{algorithm}

Then, substituting (\ref{pppppppp}) into (\ref{aaaaa}), we obtain
\begin{equation}
\|P^t(\mathbf{x}, \cdot)-\pi(\cdot)\|_{TV}\leq(1-\delta)^{t},
\label{bbbbb}
\end{equation}
completing the proof.
\end{proof}

Obviously, given the value of $\delta<1$, the mixing time of the Markov chain can be calculated by (\ref{mixing time}) and (\ref{bbbbb}), that is,
\begin{equation}
t_{\text{mix}}(\epsilon)=\frac{\text{ln}\hspace{.1em}\epsilon}{\text{ln}(1-\delta)}\leq(-\text{ln}\hspace{.1em}\epsilon)\cdot\left(\frac{1}{\delta}\right),\ \ \epsilon < 1
\label{upperboundmixing}
\end{equation}
where we use the bound $\text{ln}(1-\delta)<-\delta$ for $0<\delta<1$. Therefore, the mixing time is proportional to $1/\delta$, and becomes $O(1)$ as $\delta \to 1$.

Here, we point out that the aforementioned spectral gap $\gamma$ of the transition matrix can also be used to bound the mixing time. Resorting to the \emph{conductance} of the Markov chain \cite{mixingtimemarkovchain}, one obtains a lower bound on the spectral gap $\gamma$ of the transition matrix (see Appendix~\ref{proofl22} for its derivation)
\begin{equation}
\gamma\geq\frac{\delta^2}{8}.
\label{spectral1}
\end{equation}
Then, substituting (\ref{spectral1}) into (\ref{mixing times}) yields another upper bound on the mixing time
\begin{equation}
t_{\text{mix}}(\epsilon)\leq-\log(\pi_{\min}\epsilon)\cdot\left(\frac{8}{\delta^2}\right),\ \ \epsilon < 1,
\end{equation}
which is however looser than (\ref{upperboundmixing}).

\subsection{Convergence in General Cases ($\overline{\sigma}\neq\sigma$)}
In the proposed independent MHK algorithm, by default, the standard deviation of the proposal distribution $q$ is set the same as $\sigma$, namely, $\overline{\sigma}=\sigma$. Therefore, a natural question is whether a flexible standard deviation $\overline{\sigma}\neq\sigma$ still works. For this reason, in what follows, the relationship between $\overline{\sigma}$ and $\sigma$ is investigated.

Let the standard deviations of $q(\mathbf{x})$ and $\pi(\mathbf{x})$ be $\overline{\sigma}$ and $\sigma$ respectively, then the corresponding ratio of $q(\mathbf{x})/\pi(\mathbf{x})$ in (\ref{x1}) can be rewritten as
\begin{equation}
\ \ \ \ \ \ \ \ \ \ \ \frac{q(\mathbf{x})}{\pi(\mathbf{x})}\geq\frac{\rho_{\sigma, \mathbf{c}}(\mathbf{\Lambda})}{\prod^n_{i=1}\rho_{\overline{\sigma}_i}(\mathbb{Z})}\cdot e^{-(\frac{1}{2\overline{\sigma}^2}-\frac{1}{2\sigma^2})\|\mathbf{Bx}-\mathbf{c}\|^2}.
\end{equation}

Unfortunately,  in the case of $\overline{\sigma}<\sigma$, as $\|\mathbf{Bx}-\mathbf{c}\|$ can be arbitrary, it is impossible to determine a constant lower bound upon $q(\mathbf{x})/\pi(\mathbf{x})$ for $\mathbf{x}\in\mathbb{Z}^n$, implying the uniform ergodicity can not be achieved \cite{MengersenRate}\footnote{In theory, that $q(\mathbf{x})/\pi(\mathbf{x})$ is lower bounded by a constant for all $\mathbf{x}\in\mathbb{Z}^n$ is both sufficient and necessary to the uniform ergodicity \cite{MengersenRate}.}. Therefore, $\overline{\sigma}<\sigma$ should be avoided in practice and the corresponding convergence analysis is ignored here.

On the other hand, in the case of $\overline{\sigma}>\sigma$, let $d(\Lambda, \mathbf{c})$ denote the Euclidean distance between lattice $\Lambda$ and $\mathbf{c}$
\begin{equation}
d(\Lambda, \mathbf{c})=\underset{\mathbf{x}\in\mathbb{Z}^n}{\text{min}}\|\mathbf{Bx}-\mathbf{c}\|,
\end{equation}
then it follows that
\begin{equation}
\ \ \ \ \ \ \ \ \ \ \ \frac{q(\mathbf{x})}{\pi(\mathbf{x})}\geq\frac{\rho_{\sigma, \mathbf{c}}(\mathbf{\Lambda})}{\prod^n_{i=1}\rho_{\overline{\sigma}_i}(\mathbb{Z})}\cdot e^{-(\frac{1}{2\overline{\sigma}^2}-\frac{1}{2\sigma^2})d^2(\Lambda, \mathbf{c})}
\end{equation}
for all $\mathbf{x}\in\mathbb{Z}^n$, which means the underlying Markov chain is uniformly ergodic by satisfying (\ref{xxxx}) in Lemma 1. More precisely, $q(\mathbf{x})/\pi(\mathbf{x})$ could be expressed as
\begin{equation}
\ \ \ \ \ \ \ \ \ \ \ \frac{q(\mathbf{x})}{\pi(\mathbf{x})}\geq\frac{\rho_{\sigma, \mathbf{c}}(\mathbf{\Lambda})}{\prod^n_{i=1}\rho_{\sigma_i}(\mathbb{Z})}\cdot\beta
\label{xx1}
\end{equation}
where
\begin{equation}
\beta=\frac{\prod^n_{i=1}\rho_{\sigma_i}(\mathbb{Z})}{\prod^n_{i=1}\rho_{\overline{\sigma}_i}(\mathbb{Z})}\cdot e^{-(\frac{1}{2\overline{\sigma}^2}-\frac{1}{2\sigma^2})d(\Lambda, \mathbf{c})^2}.
\label{oooooooo}
\end{equation}
Clearly, parameter $\beta$ becomes the key to govern the convergence performance. Compared to (\ref{x1}), if $\beta>1$, the convergence of the Markov chain will be boosted by a larger value of $\delta$, otherwise the convergence will be slowed down. However, in the case of $\overline{\sigma}>\sigma$, it easy to check that the value of $\beta$ is monotonically decreasing with the given $\sigma$, rendering $\beta>1$ inapplicable to the most cases of interest.

\begin{figure}[t]
\begin{center}
\includegraphics[width=3.2in,height=2.2in]{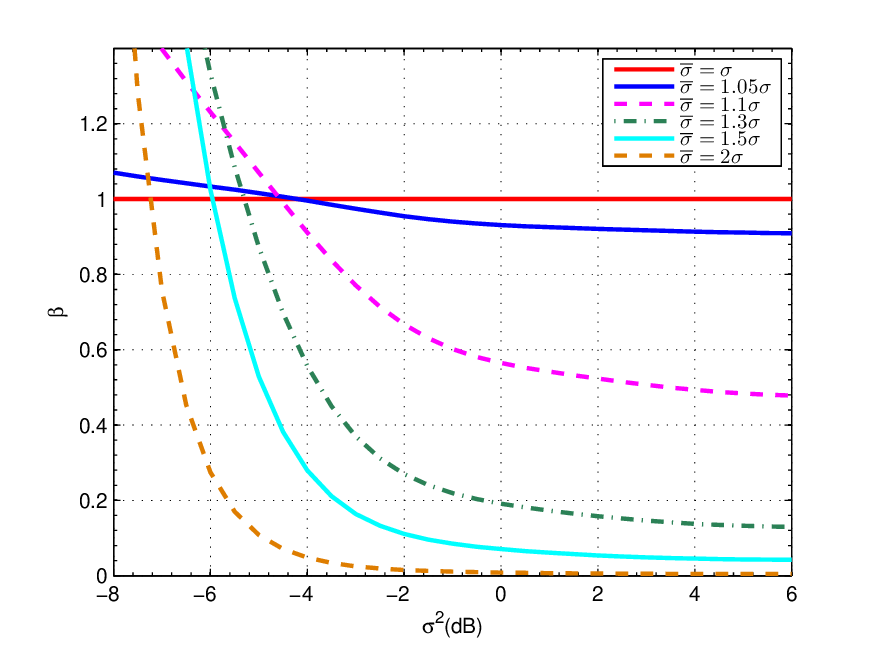}
\end{center}
\vspace{-1em}
  \caption{Coefficient $\beta$ of $E_8$ lattice in the case of $\overline{\sigma}>\sigma$ when $\mathbf{c}=\mathbf{0}$.}
  \label{simulation 5}
\end{figure}

As can be seen clearly from Fig. \ref{simulation 5}, the convergence rate can be enhanced by $\beta>1$ only for a small enough $\sigma$ (e.g., $\sigma^2<0.398$, e.g., $-4$ dB), thus making the choice of $\overline{\sigma}=\sigma$ (i.e., $\beta=1$) reasonable to maintain the convergence performance. This essentially explains the reason why the independent MHK algorithm is proposed with $\overline{\sigma}=\sigma$ as a default configuration in general.

\section{Convergence Rate Analysis}
In this section, convergence analysis about the exponential decay coefficient $\delta$ in the independent MHK algorithm is performed, which leads to a quantitative estimate of the mixing time. For a better understanding, the analysis is carried out in cases $\mathbf{c}=\mathbf{0}$ and $\mathbf{c}\neq\mathbf{0}$ separately.

\subsection{Convergence Rate ($\mathbf{c}=\mathbf{0}$)}
Lemma 1 shows that the ratio $q(\mathbf{x})/\pi(\mathbf{x})$ in the independent MHK sampling algorithm is lower bounded by a constant $\delta$. We further derive an explicit expression of the coefficient $\delta$ due to its significant impact on the convergence rate, for the case $\mathbf{c}=\mathbf{0}$.

Specifically, we have
\begin{eqnarray}
\frac{q(\mathbf{x})}{\pi(\mathbf{x})}&=&\frac{\rho_{\sigma, \mathbf{0}}(\mathbf{\Lambda})}{\prod^n_{i=1}\rho_{\sigma_i, \widetilde{x}_i}(\mathbb{Z})} \notag \\
&\overset{(c)}{\geq}&\frac{\sum_{\mathbf{x} \in \mathbb{Z}^n}e^{-\frac{1}{2\sigma^2}\parallel \mathbf{Bx} \parallel^2}}{\prod^n_{i=1}\rho_{\sigma_i}(\mathbb{Z})} \notag \\
&\overset{(d)}{=}&\frac{\Theta_{\Lambda}(\frac{1}{2\pi\sigma^2})}{\prod^n_{i=1}\Theta_{\mathbb{Z}}(\frac{1}{2\pi\sigma_i^2})} \notag \\
&\overset{(e)}{=}&\frac{\Theta_{\Lambda}(\frac{1}{s^2})}{\prod^n_{i=1}\vartheta_{3}(\frac{1}{s_i^2})}=\delta.
\label{delta1}
\end{eqnarray}
Here, for notational simplicity, $s=\sqrt{2\pi}\sigma$ and $s_i=\sqrt{2\pi}\sigma_i=s/\|\mathbf{\widehat{b}}_i\|$ are applied in the equations.
In $(c)$, the inequality $\rho_{\sigma_i, \tilde{x}}(\mathbb{Z})\leq \rho_{\sigma_i}(\mathbb{Z}) $ shown in (\ref{nnnnnn}) is used again. Theta series $\Theta_{\Lambda}$ and Jacobi theta function $\vartheta_3$ are applied in $(d)$ and $(e)$ respectively, where
\begin{equation}
\Theta_{\Lambda}(\tau)=\sum_{\mathbf{\lambda} \in \Lambda}e^{-\pi\tau\parallel \mathbf{\lambda} \parallel^2},
\end{equation}
\begin{equation}
\vartheta_3(\tau)=\sum^{+\infty}_{n=-\infty}e^{-\pi\tau n^2}
\end{equation}
with $\Theta_{\mathbb{Z}}=\vartheta_3$ \cite{ConwayandSloane}.

\begin{my4}
If $s\geq \sqrt{\omega(\log n)}\cdot\max_{1\leq i \leq n}\|\mathbf{\widehat{b}}_i\|$ or $s\leq \sqrt{\omega(\log n)}^{-1}\cdot\min_{1\leq i \leq n}\|\widehat{\mathbf{b}}_{i}\|$, then the coefficient $\delta\approx 1$.
\end{my4}
\begin{proof}
To start with, let us recall the \emph{flatness factor} \cite{LLBS_12}, which is defined as
\begin{equation}
\epsilon_{\Lambda}(\sigma)=\frac{\det(\mathbf{B})}{(\sqrt{2\pi}\sigma)^n}\Theta_{\Lambda}\left(\frac{1}{2\pi\sigma^2}\right)-1.
\label{flatness1}
\end{equation}
and
\begin{equation}
\epsilon_{\Lambda}(\sigma)=\varepsilon,\ \ \text{if}\ \ \sigma=\eta_{\varepsilon}(\Lambda).
\label{delta2}
\end{equation}
Here, $\eta_{\varepsilon}(\Lambda)$ is known as the \emph{smoothing parameter} and for any $n$-dimensional lattice $\Lambda$ and positive real $\varepsilon>0$, $\eta_{\varepsilon}(\Lambda)$ is defined as the smallest real $\sigma>0$ such that $\rho_{1/\sqrt{2\pi}\sigma}(\Lambda^{*}\backslash\{\mathbf{0}\})\leq\varepsilon$, where $\Lambda^{*}$ denotes the dual lattice of $\Lambda$ \cite{Trapdoor}.

Therefore, the exponential decay coefficient $\delta$ given in (\ref{delta1}) can be expressed as
\begin{eqnarray}
\delta&=&\frac{\Theta_{\Lambda}(\frac{1}{2\pi\sigma^2})}{\prod^n_{i=1}\vartheta_{3}(\frac{1}{2\pi\sigma_i^2})}\notag\\
&=&\frac{|\det(\mathbf{B})|^{-1}\cdot(\sqrt{2\pi}\sigma)^{n}\cdot[\epsilon_{\Lambda}(\sigma)+1]}{\prod^n_{i=1}\sqrt{2\pi}\sigma_i\cdot[\epsilon_{\mathbb{Z}}(\sigma_i)+1]}\notag\\
&=&\frac{\epsilon_{\Lambda}(\sigma)+1}{\prod^n_{i=1}[\epsilon_{\mathbb{Z}}(\sigma_i)+1]},
\end{eqnarray}
where $\text{det}(\cdot)$ denotes the determinant of a matrix.

Meanwhile, from \cite[Lemma 3.3]{MicciancioGaussian}, for any $n$-dimensional lattice $\Lambda$ and positive real $\varepsilon>0$, it follows that
\begin{equation}
\eta_{\varepsilon}(\Lambda)\leq \sqrt{\frac{\log(2n(1+1/\varepsilon))}{\pi}}\cdot\max_{1\leq i \leq n}\|\mathbf{\widehat{b}}_i\|
\end{equation}
and for any $\omega(\log n)$, there is a negligible $\varepsilon(n)$ such that
\begin{equation}
\eta_{\varepsilon}(\Lambda)\leq \sqrt{\omega(\log n)}\cdot\max_{1\leq i \leq n}\|\mathbf{\widehat{b}}_i\|.
\end{equation}
According to (\ref{flatness1}), it is easy to verify that the flatness factor $\epsilon_{\Lambda}(\sigma)$ is a monotonically decreasing function of $\sigma$, i.e., for $\sigma_1\geq\sigma_2$, we have $\epsilon_{\Lambda}(\sigma_1)\leq\epsilon_{\Lambda}(\sigma_2)$. Therefore, letting $\eta_{\varepsilon}(\Lambda)\leq \sqrt{\omega(\log n)}\cdot\max_{1\leq i \leq n}\|\mathbf{\widehat{b}}_i\|$ be a benchmark of comparison, we may bound the flatness factor $\epsilon_{\Lambda}(\sigma)$ by a negligible $\varepsilon(n)$ if $\sigma>\sqrt{\omega(\log n)}\cdot\max_{1\leq i \leq n}\|\mathbf{\widehat{b}}_i\|$. On the other hand, it is also easy to check that $\epsilon_{\mathbb{Z}}(\sigma_i)$ will become negligible if $\sigma>\sqrt{\omega(\log n)}\cdot\max_{1\leq i \leq n}\|\mathbf{\widehat{b}}_i\|$. Hence, we have
\begin{equation}
\delta=\frac{\epsilon_{\Lambda}(\sigma)+1}{\prod^n_{i=1}[\epsilon_{\mathbb{Z}}(\sigma_i)+1]}\approx1
\end{equation}
for $\sigma>\sqrt{\omega(\log n)}\cdot\max_{1\leq i \leq n}\|\mathbf{\widehat{b}}_i\|$.

On the other hand, according to \emph{Jacobi's formula} \cite{BelfioreConstructionandAnalysis}
\begin{equation}
\Theta_{\Lambda}(\tau)=|\text{det}(\mathbf{B})|^{-1}\left(\frac{1}{\tau}\right)^{\frac{n}{2}}\Theta_{\Lambda^{*}}\left(\frac{1}{\tau}\right),
\label{Jacobi's formula}
\end{equation}
the expression of the flatness factor shown in (\ref{flatness1}) can be rewritten as
\begin{equation}
\epsilon_{\Lambda}(\sigma)=\Theta_{\Lambda^{*}}(2\pi\sigma^2)-1,
\end{equation}
where $\Lambda^{*}$ is the dual lattice of $\Lambda$.
Then, we have
\begin{eqnarray}
\delta&=&\frac{\Theta_{\Lambda}(\frac{1}{2\pi\sigma^2})}{\prod^n_{i=1}\vartheta_{3}(\frac{1}{2\pi\sigma_i^2})}\notag\\
&=&\frac{\epsilon_{\Lambda^{*}}(\frac{1}{2\pi\sigma})+1}{\prod^n_{i=1}[\epsilon_{\mathbb{Z^{*}}}(\frac{1}{2\pi\sigma_i})+1]},
\label{mnmn}
\end{eqnarray}
where $\mathbb{Z}^{*}=\mathbb{Z}$.

With respect to $\epsilon_{\Lambda^{*}}(\frac{1}{2\pi\sigma})$ and $\epsilon_{\mathbb{Z}^{*}}(\frac{1}{2\pi\sigma_i})$ in (\ref{mnmn}), similarly, if
\begin{equation}
\frac{1}{2\pi\sigma}\geq \sqrt{\omega(\log n)}\cdot\max_{1\leq i \leq n}\|\widehat{\mathbf{b}}^{*}_i\|,
\label{sssss}
\end{equation}
where $\widehat{\mathbf{b}}^{*}_i$'s are the Gram-Schmidt vectors of the dual lattice basis $\mathbf{B}^{*}$,
then both $\epsilon_{\Lambda^{*}}(\frac{1}{2\pi\sigma})$ and $\epsilon_{\mathbb{Z}^{*}}(\frac{1}{2\pi\sigma_i})$ will be bounded by a negligible $\varepsilon(n)$. Thus, we have
\begin{equation}
\delta\approx1.
\end{equation}

According to (\ref{sssss}), it follows that
\begin{eqnarray}
\sigma&\leq& \sqrt{\omega(\log n)}^{-1}\cdot\left(\max_{1\leq i \leq n}\|\widehat{\mathbf{b}}^{*}_i\|\right)^{-1}\notag\\
&\overset{(f)}{=}&\sqrt{\omega(\log n)}^{-1}\cdot\left[\max_{1\leq i \leq n}(\|\widehat{\mathbf{b}}_{n-i+1}\|^{-1})\right]^{-1}\notag\\
&=&\sqrt{\omega(\log n)}^{-1}\cdot\left[\left(\min_{1\leq i \leq n}\|\widehat{\mathbf{b}}_{i}\|\right)^{-1}\right]^{-1}\notag\\
&=&\sqrt{\omega(\log n)}^{-1}\cdot\min_{1\leq i \leq n}\|\widehat{\mathbf{b}}_{i}\|,
\end{eqnarray}
where $(f)$ comes from the fact that \cite{CongProxity}
\begin{equation}
\|\widehat{\mathbf{b}}^{*}_i\|=\|\widehat{\mathbf{b}}_{n-i+1}\|^{-1}.
\end{equation}
Therefore, the proof is completed.
\end{proof}

Obviously, according to Proposition 1, as $s$ either goes to $0$ or $\infty$, the coefficient $\delta$ will converge to 1. We remark that this is in line with the fact that Klein's algorithm is capable of sampling from the lattice Gaussian distribution directly when $\sigma>\sqrt{\omega(\log n)}\cdot\max_{1\leq i \leq n}\|\mathbf{\widehat{b}}_i\|$.
\begin{my4}
If $s\leq\min_{1\leq i\leq n}\|\widehat{\mathbf{b}}_i\|$, then
the coefficient $\delta$ is lower bounded by
\begin{equation}
\delta\geq1.086^{-n}\cdot\Theta_{\Lambda}(\frac{1}{s^2}).
\end{equation}
Meanwhile, if $s\geq\max_{1\leq i\leq n}\|\widehat{\mathbf{b}}_i\|$, then
the coefficient $\delta$ is lower bounded by
\begin{equation}
\delta\geq1.086^{-n}\cdot\Theta_{\Lambda^{*}}(s^2).
\end{equation}
\end{my4}
\renewcommand{\arraystretch}{1.8}
\begin{table}[t]
\begin{center}
\caption{Lower bounds on $\delta$ with respect to $s=\sqrt{2\pi}\sigma$ in the independent MHK algorithm.}
\label{tab:TCQvsPL}
\resizebox{\columnwidth}{!}{%
\begin{tabular}{|c||c|}\hline
$s\hspace{-0.2em}\leq \hspace{-0.2em}[\sqrt{2\pi\omega(\log n)}]^{-1}\hspace{-0.2em}\cdot\hspace{-0.3em}\underset{1\leq i \leq n}{\min}\|\widehat{\mathbf{b}}_{i}\|$  &  $\delta\approx1$   \\\hline
$s\leq\underset{1\leq i \leq n}{\min}\|\widehat{\mathbf{b}}_i\|$  & $\delta\geq1.086^{-n}\cdot\Theta_{\Lambda}(\frac{1}{s^2})$    \\\hline
$\underset{1\leq i \leq n}{\min}\|\widehat{\mathbf{b}}_i\|\hspace{-0.2em}\leq \hspace{-0.2em}s\hspace{-0.2em}\leq\hspace{-0.2em}\underset{1\leq i \leq n}{\max}\|\widehat{\mathbf{b}}_i\|$  &  $\delta\hspace{-0.2em}\geq\hspace{-0.2em}1.086^{-(n-m)}\hspace{-0.2em}\cdot\hspace{-0.2em}2^{-m}\hspace{-0.2em}\cdot\hspace{-0.2em}\frac{\prod_{i\in I}\|\widehat{\mathbf{b}}_i\|}{s^m}\hspace{-0.2em}\cdot\hspace{-0.2em}\Theta_{\Lambda}(\hspace{-0.1em}\frac{1}{s^2}\hspace{-0.1em})$  \\\hline
$s\geq\underset{1\leq i \leq n}{\max}\|\widehat{\mathbf{b}}_i\|$  &  $\delta\geq1.086^{-n}\cdot\Theta_{\Lambda^{*}}(s^2)$  \\\hline
$s\geq \sqrt{2\pi\omega(\log n)}\cdot\underset{1\leq i \leq n}{\max}\|\mathbf{\widehat{b}}_i\|$  &  $\delta\approx 1$ \\\hline
\end{tabular}
}
\end{center}
\end{table}
\begin{proof}
By definition, we have
\begin{equation}
\vartheta_3(1)=\sum^{+\infty}_{n=-\infty}e^{-\pi n^2}=\frac{\sqrt[4]{\pi}}{\Gamma(\frac{3}{4})}=1.086,
\end{equation}
where $\Gamma(\cdot)$ stands for the Gamma function \cite{ThetaExplicit}. It is worth pointing out that the explicit values of $\vartheta_3(2)$, $\vartheta_3(3)$, \ldots can also be calculated \cite{ThetaBook}, where the same derivation in the following can also be carried out. Here we choose $\vartheta_3(1)$ as the benchmark due to its simplicity. As the Jacobi theta function $\vartheta_3(\tau)$ is monotonically decreasing with $\tau$, let $1/s^2_i\geq1$, i.e., $s\leq\|\widehat{\mathbf{b}}_i\|$, then it follows that
\begin{equation}
\vartheta_{3}(\frac{1}{s_i^2})\leq \vartheta_{3}(1)=1.086.
\label{p111}
\end{equation}
Assume $s\leq\min_{1\leq i\leq n}\|\widehat{\mathbf{b}}_i\|$, then the following lower bound for $\delta$ can be obtained,
\begin{equation}
\delta=\frac{\Theta_{\Lambda}(\frac{1}{s^2})}{\prod^n_{i=1}\vartheta_{3}(\frac{1}{s_i^2})}\geq1.086^{-n}\cdot\Theta_{\Lambda}(\frac{1}{s^2}).
\end{equation}

On the other hand, as $\mathbb{Z}$ is a self-dual lattice, i.e., $\mathbb{Z}=\mathbb{Z}^{*}$, then if $s^2_i\geq1$, namely, $s\geq\|\widehat{\mathbf{b}}_i\|$, it follows that
\begin{equation}
\vartheta_{3}^{*}(s_i^2)=\vartheta_{3}(s_i^2)\leq \vartheta_{3}(1)\leq1.086.
\end{equation}
Therefore, let $s\geq\max_{1\leq i\leq n}\|\widehat{\mathbf{b}}_i\|$, according to Jacobi's formula shown in (\ref{Jacobi's formula}), $\delta$ can be lower bounded as
{\allowdisplaybreaks\begin{eqnarray}
\delta&=&\frac{\Theta_{\Lambda}(\frac{1}{s^2})}{\prod^n_{i=1}\vartheta_{3}(\frac{1}{s_i^2})}\notag\\
&=&\frac{|\text{det}(\mathbf{B})|^{-1}(s^2)^{\frac{n}{2}}\Theta_{\Lambda^{*}}(s^2)}{\prod^n_{i=1}(s_i^2)^{\frac{n}{2}}\vartheta_3^{*}(s_i^2)}\notag\\
&=&\frac{\Theta_{\Lambda^{*}}(s^2)}{\prod^n_{i=1}\vartheta_3^{*}(s_i^2)}\notag\\
&\geq&1.086^{-n}\cdot\Theta_{\Lambda^{*}}(s^2),
\end{eqnarray}}
completing the proof.
\end{proof}
\emph{Remark:} We emphasize that the significance of lattice reduction (e.g., LLL or HKZ) can be seen here, as increasing $\min_{1\leq i\leq n}\|\widehat{\mathbf{b}}_i\|$ and decreasing $\max_{1\leq i\leq n}\|\widehat{\mathbf{b}}_i\|$ simultaneously will greatly enhance the convergence performance due to a better lower bound of $\delta$.

Next, with respect to the range of $\min_{1\leq i\leq n}\|\widehat{\mathbf{b}}_i\|\leq s\leq\max_{1\leq i\leq n}\|\widehat{\mathbf{b}}_i\|$, we arrive at the following proposition.
\begin{my4}
If $\min_{1\leq i\leq n}\|\widehat{\mathbf{b}}_i\|\leq s\leq\max_{1\leq i\leq n}\|\widehat{\mathbf{b}}_i\|$, then the coefficient $\delta$ is lower bounded by
\begin{equation}
\delta\geq1.086^{-(n-m)}\cdot2^{-m}\cdot\frac{\prod_{i\in I}\|\widehat{\mathbf{b}}_i\|}{s^m}\cdot\Theta_{\Lambda}\left(\frac{1}{s^2}\right),
\end{equation}
where $I$ denotes the subset of indexes $i$ with $s_i>1$ (i.e., $s>\|\widehat{\mathbf{b}}_i\|$), $i\in \{1, 2, \ldots, n\}$, $|I|=m$.
\end{my4}
\begin{proof}
From the definition, we have
\begin{eqnarray}
\vartheta_3(\tau)&=&\sum^{+\infty}_{n=-\infty}e^{-\pi\tau n^2}\notag\\
&=&1+2\sum_{n\geq1}e^{-\pi\tau n^2}\notag\\
&\leq&1+2\int^{\infty}_0e^{-\pi\tau x^2}dx\notag\\
&\overset{(g)}{=}&1+\sqrt{\frac{1}{\tau}},
\end{eqnarray}
where $(g)$ holds due to the \emph{Gaussian integral} $\int^{\infty}_{-\infty}e^{-ax^2}dx=\sqrt{\frac{\pi}{a}}$.
\begin{figure}[t]
\begin{center}
\includegraphics[width=3.2in,height=2.2in]{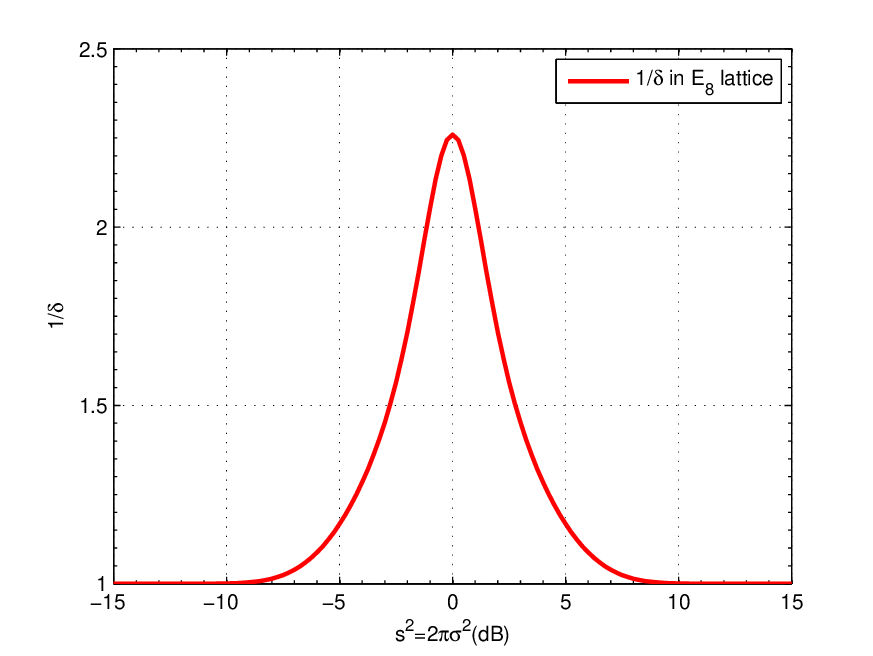}
\end{center}
\vspace{-1em}
  \caption{Coefficient $1/\delta$ of the $E_8$ lattice in the case of $\mathbf{c}=\mathbf{0}$.}
  \label{simulation 1}
\end{figure}

Hence, for terms $\vartheta_3(\frac{1}{s_i^2})$ with $1/s^2_i\leq1$, namely, $s\geq\|\widehat{\mathbf{b}}_i\|$, we have
\begin{equation}
\vartheta_{3}\left(\frac{1}{s_i^2}\right)\leq 1+|s_i|\leq2s_i=2\frac{s}{\|\widehat{\mathbf{b}}_i\|}.
\label{p222}
\end{equation}
Therefore, from (\ref{p111}) and (\ref{p222}), if follows that
\begin{equation}
\prod^n_{i=1}\vartheta_{3}\left(\frac{1}{s_i^2}\right)\leq1.086^{(n-m)}\cdot2^{m}\cdot\frac{s^m}{\prod_{i\in I}\|\widehat{\mathbf{b}}_i\|},
\end{equation}
completing the proof.
\end{proof}

To summarize, the value of $\delta$ with respect to the given $s=\sqrt{2\pi}\sigma$ in the independent MHK algorithm is given in Table I.

Now, let us consider some lattices whose theta series are more understood. We have the following property for an isodual lattice, which is one that is geometrically similar to its dual \cite{BelfioreConstructionandAnalysis}.
\begin{my4}
The coefficient $\delta=\frac{\Theta_{\Lambda}(\frac{1}{s^2})}{\prod^n_{i=1}\vartheta_{3}(\frac{1}{s_i^2})}$ for an isodual lattice $\Lambda$ has a multiplicative symmetry point at $s = 1$, and asymptotically converges to 1 on both sides when $s$ either goes to $0$ or $\infty$.
\end{my4}

\begin{proof}
Here, we note that the theta series $\Theta_{\Lambda}$ of an isodual lattice $\Lambda$ and that of its dual $\Lambda^{*}$ are the same, i.e., $\Theta_{\Lambda}(\tau)=\Theta_{\Lambda^{*}}(\tau)$, and the volume of
an isodual lattice $|\text{det}(\mathbf{B})|$ naturally equals $1$. Therefore, we have
\begin{eqnarray}
\Theta_{\Lambda}\left(\frac{1}{s^2}\right)&=&s^n\Theta_{\Lambda}(s^2),
\label{theta Lamda s}
\end{eqnarray}
\vspace{-1em}
\begin{eqnarray}
\vartheta_3\left(\frac{1}{s_i^2}\right)&=&s_i\vartheta_3(s_i^2),
\label{theta Z s}
\end{eqnarray}
then from (\ref{theta Lamda s}) and (\ref{theta Z s}), the symmetry with respect to $s=1$ can be obtained as follows,
\begin{eqnarray}
\frac{\Theta_{\Lambda}(\frac{1}{s^2})}{\prod^n_{i=1}\vartheta_{3}(\frac{1}{s_i^2})}&=&\frac{s^n\Theta_{\Lambda}(s^2)}{\prod^n_{i=1}s_i\vartheta_3(s_i^2)}\notag\\
&=&\frac{\Theta_{\Lambda}(s^2)}{\prod^n_{i=1}\frac{1}{\|\mathbf{\widehat{b}}_i\|}\vartheta_3(s_i^2)}\notag\\
&=&\frac{\Theta_{\Lambda}(s^2)}{\frac{1}{|\text{det}(\mathbf{B})|}\cdot\prod^n_{i=1}\vartheta_3(s_i^2)}\notag\\
&=&\frac{\Theta_{\Lambda}(s^2)}{\prod^n_{i=1}\vartheta_3(s_i^2)}.
\end{eqnarray}

By definition, it is straightforward to verify that
\begin{equation}
\frac{\Theta_{\Lambda}(\frac{1}{s^2})}{\prod^n_{i=1}\vartheta_{3}(\frac{1}{s_i^2})}\rightarrow1,\  \text{when} \ \ s\rightarrow0.
\end{equation}
Then because of the symmetry, $\frac{\Theta_{\Lambda}(\frac{1}{s^2})}{\prod^n_{i=1}\vartheta_{3}(\frac{1}{s_i^2})}$ will also
asymptotically approach $1$ when $s \to \infty$, completing the proof.
\end{proof}

\begin{figure}[t]
\begin{center}
\includegraphics[width=3.2in,height=2.2in]{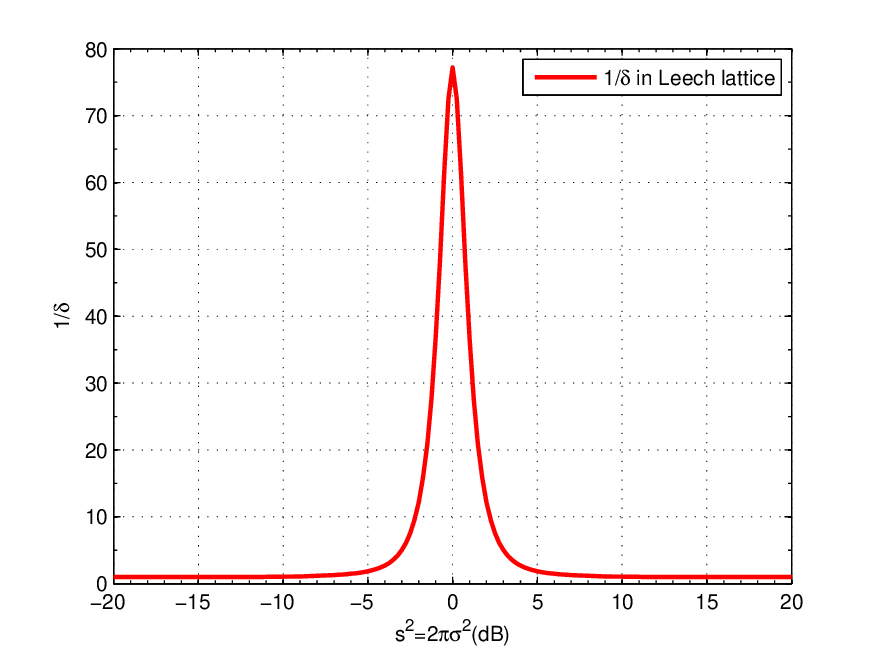}
\end{center}
\vspace{-1em}
  \caption{Coefficient $1/\delta$ of the Leech lattice in the case of $\mathbf{c}=\mathbf{0}$.}
  \label{simulation 2}
\end{figure}

\begin{figure}[t]
\begin{center}
\includegraphics[width=3.2in,height=2.2in]{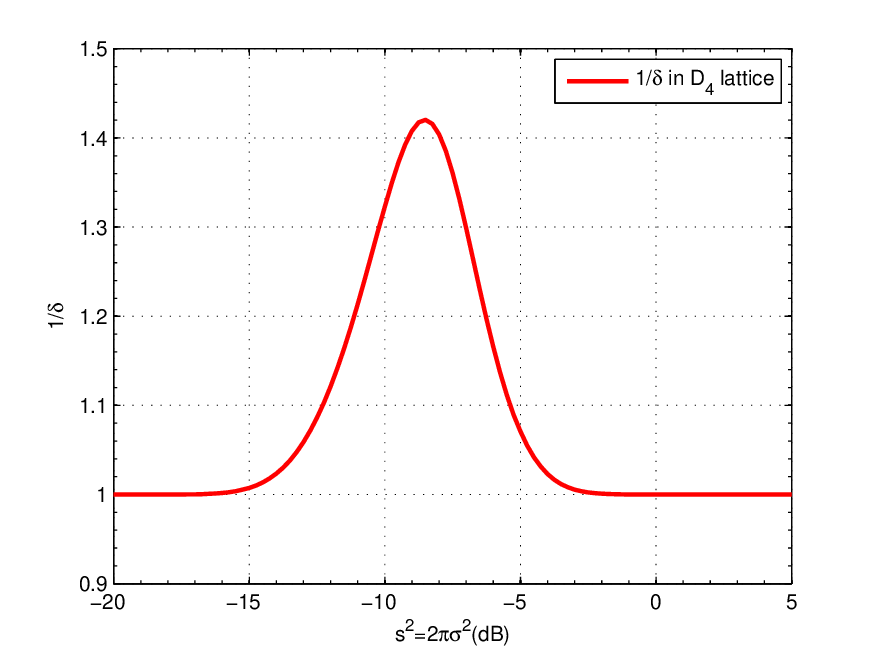}
\end{center}
\vspace{-1em}
  \caption{Coefficient $1/\delta$ of the $D_4$ lattice in the case of $\mathbf{c}=\mathbf{0}$.}
  \label{simulation 3}
\end{figure}

Examples of the coefficient $1/\delta$ for the isodual $E_8$ and Leech lattice are shown in Fig. \ref{simulation 1} and Fig. \ref{simulation 2}, respectively. It is worth pointing out that $1/\delta$ has a maximum at the symmetry point $s=1$, i.e., $\sigma^2=\frac{1}{2\pi}$. Actually, $1/\delta$ is similar to, but not exactly the same as the \emph{secrecy gain} defined in \cite{BelfioreConstructionandAnalysis}. In our context, $1/\delta$ roughly estimates the number of the Markov moves required to reach the stationary distribution. On the other hand, as for non-isodual lattices, $D_4$ lattice is applied to give the illustration in Fig. \ref{simulation 3}, where the symmetry still holds but centers at $s=0.376$. Therefore, with the exact value of $\delta$, the explicit estimation of the mixing time for the underlying Markov chain can be obtained.

\subsection{Convergence Rate ($\mathbf{c}\neq \mathbf{0}$)}
As for the convergence analysis in the case of $\mathbf{c}\neq \mathbf{0}$, we firstly define the exponential decay coefficient $\delta'$ as
\begin{equation}
\delta'=\frac{q(\mathbf{x})}{\pi(\mathbf{x})}=\frac{\rho_{\sigma, \mathbf{c}}(\mathbf{\Lambda})}{\prod^n_{i=1}\rho_{\sigma_i, \widetilde{x}_i}(\mathbb{Z})},
\end{equation}
then we have the following proposition.
\begin{my4}
For any $\mathbf{c}\in \mathbb{R}^n$ and $\mathbf{c}\neq\mathbf{0}$, one has
\begin{equation}
\delta'\geq e^{-\frac{d^2(\Lambda, \mathbf{c})}{2\sigma^2}}\cdot\delta.
\label{mmm1}
\end{equation}
\end{my4}
\begin{proof}
Let $\mathbf{c}'=\mathbf{c}\mod \Lambda$ stand for the modular operation of $\mathbf{c}$ over lattice $\Lambda$. Then it follows that
\begin{eqnarray}
\rho_{\sigma, \mathbf{c}}(\mathbf{\Lambda})\hspace{-.9em}&=&\hspace{-.8em}\sum_{\mathbf{z} \in \mathbf{\Lambda} }e^{-\frac{1}{2\sigma^2}\parallel \mathbf{z}-\mathbf{c} \parallel^2} \notag\\
&=&\hspace{-.8em}\sum_{\mathbf{z} \in \mathbf{\Lambda} }e^{-\frac{1}{2\sigma^2}\parallel \mathbf{z}-\mathbf{c}' \parallel^2} \notag\\
\hspace{-.9em}&=&\hspace{-.8em}e^{-\frac{\|\mathbf{c}'\|^2}{2\sigma^2}}\hspace{-.2em}\cdot\hspace{-.2em}\sum_{\mathbf{z} \in \mathbf{\Lambda}
}\hspace{-.1em}e^{-\frac{\|\mathbf{z}\|^2}{2\sigma^2}\hspace{-.1em}}\hspace{-.4em}\cdot\hspace{-.2em}\frac{1}{2}\hspace{-.2em}\cdot\hspace{-.2em}\left(\hspace{-.2em}e^{-\frac{1}{\sigma^2}\hspace{-.1em}\langle\mathbf{z}, \mathbf{c}'\rangle}\hspace{-.3em}+\hspace{-.2em}e^{\frac{1}{\sigma^2}\hspace{-.1em}\langle\mathbf{z}, \mathbf{c}'\rangle}\hspace{-.3em}\right)\notag\\
\hspace{-.9em}&\overset{(h)}{\geq}&\hspace{-.8em}e^{-\frac{\|\mathbf{c}'\|^2}{2\sigma^2}}\cdot\sum_{\mathbf{z} \in \mathbf{\Lambda}}e^{-\frac{\|\mathbf{z}\|}{2\sigma^2}} \notag\\
\hspace{-.9em}&=&\hspace{-.8em}e^{-\frac{d^2(\Lambda, \mathbf{c})}{2\sigma^2}}\cdot\rho_{\sigma}(\mathbf{\Lambda}),
\label{jjjjj}
\end{eqnarray}
where $(h)$ follows from the fact that for any positive real $a>0$, $a+1/a\geq2$.
\end{proof}
Thus, the value of $\delta'$ is reduced by a factor of $e^{-\frac{d^2(\Lambda, \mathbf{c})}{2\sigma^2}}$ from $\delta$. Clearly, if $\mathbf{c}=\mathbf{0}$, then $\delta'=\delta$, implying $\mathbf{c}\neq\mathbf{0}$ is a general case of $\mathbf{c}=\mathbf{0}$ \footnote{In fact, as $\rho_{\sigma, \mathbf{c}}(\mathbf{\Lambda})$ is periodic, all $\mathbf{c}\in \Lambda$ will lead to $d(\Lambda, \mathbf{c})=\mathbf{0}$, thus corresponding to the case of $\mathbf{c}=\mathbf{0}$.}. Hence, according to (\ref{jjjjj}), as long as $\mathbf{c}$ is not too far from $\Lambda$, $\delta'$ has a similar lower bound.

\section{Symmetric Metropolis-Klein Algorithm}
In this section, we propose the symmetrical Metropolis-Klein (SMK) algorithm for lattice Gaussian sampling. The underlying Markov chain is proved to be geometrically ergodic, which not only converges exponentially fast, but also depends on the selection of the initial state.

\subsection{Symmetric Metropolis-Klein Algorithm}
The Metropolis algorithm can be viewed as a special case of the MH algorithm by utilizing a symmetric
proposal distribution $q(\mathbf{x},\mathbf{y})=q(\mathbf{y},\mathbf{x})$ \cite{MetropolisOrignial}. In the proposed algorithm, we again use Klein's algorithm to generate the symmetric proposal distribution. Yet, the generation of the state candidate $\mathbf{y}$ depends on the current state $\mathbf{x}$, which is different from the independent MHK algorithm. Specifically, as shown in Algorithm 3, its sampling procedure at each Markov move can be summarized by the following steps:

1)\ \ \hspace{-.2em}\emph{Given the current Markov state $\mathbf{X}_{t}=\mathbf{x}$, sample from the symmetric proposal distribution through Klein's algorithm to obtain the candidate state $\mathbf{y}$ for $\mathbf{X}_{t+1}$},
\begin{equation}
q(\mathbf{x},\hspace{-.1em}\mathbf{y})\hspace{-.2em}=\hspace{-.2em}\frac{\rho_{\sigma, \mathbf{Bx}}(\mathbf{By})}{\prod^n_{i=1}\rho_{\sigma_i, \widetilde{y}_i}(\mathbb{Z})}\hspace{-.2em}=\hspace{-.2em}\frac{e^{-\frac{1}{2\sigma^2}\|\mathbf{Bx}-\mathbf{By}\|^2}}{\prod^n_{i=1}\rho_{\sigma_i, \widetilde{y}_i}(\mathbb{Z})}\hspace{-.2em}\overset{(i)}{=}\hspace{-.2em}q(\mathbf{y},\hspace{-.1em}\mathbf{x}),
\label{metropolis-klein sampling probability}
\end{equation}
where $\widetilde{y}_i=\frac{c'_i-\sum^n_{j=i+1}r_{i,j}y_j}{r_{i,i}}$, $\mathbf{c'}=\mathbf{Q}^{\dag}\mathbf{Bx}$ and $\mathbf{B}=\mathbf{QR}$. Note that equality $(i)$ holds due to the inherent symmetry (see Lemma 2 in the following).

2)\ \ \hspace{-.2em}\emph{Calculate the acceptance ratio $\alpha(\mathbf{x},\mathbf{y})$}
\begin{eqnarray}
\alpha(\mathbf{x},\mathbf{y})&=&\text{min}\left\{1,\frac{\pi(\mathbf{y})q(\mathbf{y},\mathbf{x})}{\pi(\mathbf{x})q(\mathbf{x},\mathbf{y})}\right\}=\text{min}\left\{1,\frac{\pi(\mathbf{y})}{\pi(\mathbf{x})}\right\}\notag\\
&=&\text{min}\left\{1, e^{\frac{1}{2\sigma^2}\left(\|\mathbf{Bx}-\mathbf{c}\|^2-\|\mathbf{By}-\mathbf{c}\|^2\right)}\right\},
\label{MK-acceptance probability}
\end{eqnarray}
where $\pi=D_{\Lambda,\sigma,\mathbf{c}}$.

3)\ \ \hspace{-.2em}\emph{Make a decision for $\mathbf{X}_{t+1}$ based on $\alpha(\mathbf{x},\mathbf{y})$ to accept $\mathbf{X}_{t+1}=\mathbf{y}$ or not.}

\begin{my1}
The proposal distribution $q$ shown in (\ref{metropolis-klein sampling probability}) is symmetric and only depends on $\mathbf{x}-\mathbf{y}$, namely,
\begin{equation}
q(\mathbf{x},\mathbf{y})=q(\mathbf{y},\mathbf{x})=q(\mathbf{x}-\mathbf{y})
\end{equation}
for all $\mathbf{x}, \mathbf{y}\in \mathbb{Z}^n$.
\end{my1}
The proof of Lemma 2 is provided in Appendix~\ref{proofl2}. Such a special case is called the ``random-walk" Metropolis-Hastings algorithm \cite{RobertsGeneralstatespace}.

At each Markov move, the state candidate $\mathbf{y}$ for $\mathbf{X}_{t+1}$ is sampled from a Gaussian-like distribution centered at the current state $\mathbf{x}$. Since the chain is symmetric, the calculation of the acceptance ratio $\alpha$ is greatly simplified. From (\ref{MK-acceptance probability}), it is quite
straightforward to see that if $\mathbf{By}$ is closer to the given point $\mathbf{c}$ than $\mathbf{Bx}$,
then state candidate $\mathbf{y}$ must be accepted by $\mathbf{X}^{t+1}$ since $\alpha=1$; otherwise it will be accepted with a probability depending on the distance from $\mathbf{By}$ to $\mathbf{c}$,
thus forming a Markov chain\footnote{A query about the SMK algorithm is whether a flexible standard deviation $\overline{\sigma}$ in the proposal distribution $q$ works, i.e., $\overline{\sigma}\neq\sigma$. The answer is yes. However, since the explicit convergence rate is tedious to analyze, we omit its analysis here.}.
\begin{algorithm}[t]
\caption{Symmetric Metropolis-Klein Algorithm for Lattice Gaussian Sampling}
\begin{algorithmic}[1]
\Require
$\mathbf{B}, \sigma, \mathbf{c}, \mathbf{X}_0, t_{\text{mix}}(\epsilon)$
\Ensure sample from a distribution statistically close to $\pi=D_{\Lambda,\sigma,\mathbf{c}}$
\For {$t=$1,2,\ \ldots,\ }
\State let $\mathbf{x}$ denote the state of $\mathbf{X}_{t-1}$
\State generate $\mathbf{y}$ by the proposal distribution $q(\mathbf{x},\mathbf{y})$ in (\ref{metropolis-klein sampling probability})
\State calculate the acceptance ratio $\alpha(\mathbf{x},\mathbf{y})$ in (\ref{MK-acceptance probability})
\State generate a sample $u$ from the uniform density $U[0,1]$
\If {$u\leq \alpha(\mathbf{x},\mathbf{y})$}
\State let $\mathbf{X}_t=\mathbf{y}$
\Else
\State $\mathbf{X}_t=\mathbf{x}$
\EndIf
\If {$t\geq t_{\text{mix}}(\epsilon)$}
\State output the state of $\mathbf{X}_t$
\EndIf
\EndFor
\end{algorithmic}
\end{algorithm}

Again, we recall the following standard result (see, e.g., \cite{mixingtimemarkovchain} for a proof).

\begin{my4}
Given the target lattice Gaussian distribution $\pi=D_{\Lambda,\sigma,\mathbf{c}}$, the Markov chain induced by the proposed symmetric Metropolis-Klein algorithm is ergodic:
\end{my4}
\vspace{-1em}
\begin{equation}
\underset{t\rightarrow\infty}{{\lim}}\|P^t(\mathbf{x}; \cdot)-D_{\Lambda,\sigma,\mathbf{c}}(\cdot)\|_{TV}=0
\end{equation}
\emph{for all states $\mathbf{x}\in\mathbb{Z}^n$.}

\subsection{Geometric Ergodicity}
In MCMC, a set $C\subseteq\Omega$ is referred to as a \emph{small set}, if
there exist $k>0$, $1>\delta>0$ and a probability measure $v$ on $\Omega$ such that
\begin{equation}
P^k(\mathbf{x},\mathcal{B})\geq\delta v(\mathcal{B}), \ \ \forall\mathbf{x}\in C
\label{smallset}
\end{equation}
for all measurable subsets $\mathcal{B}\subseteq \Omega$. This is also known as the \emph{minorisation condition} in literature \cite{Meynbook}. Actually, uniform ergodicity is a special case where the minorisation condition is satisfied with $C=\Omega$. For a bounded small set $C$, the \emph{drift condition} of discrete state space Markov chains is defined as follows \cite{RobertsGeneralstatespace}:

\begin{my3}
A Markov chain with discrete state space $\Omega$ satisfies the drift condition if there are constants $0<\lambda<1$ and $b<\infty$, and a
function $V:\Omega\rightarrow[1, \infty)$, such that
\begin{equation}
\sum_{\mathbf{y}\in\Omega}P(\mathbf{x},\mathbf{y})V(\mathbf{y})\leq\lambda V(\mathbf{x})+b\mathbf{1}_C(\mathbf{x})
\label{drift5}
\end{equation}
for all $\mathbf{x}\in\Omega$, where $C\subseteq\Omega$ is a small set, and the indicator function $\mathbf{1}_C(\mathbf{x})=1$ if $\mathbf{x}\in C$ and $0$ otherwise.
\end{my3}

\newcounter{TempEqCnt}                         
\setcounter{TempEqCnt}{\value{equation}} 
\setcounter{equation}{77}
\begin{figure*}[bb]
\hrulefill
\begin{eqnarray}
\sum_{\mathbf{y}\in\mathbb{Z}^n}P(\mathbf{x},\mathbf{y})V(\mathbf{y})&=&\sum_{\mathbf{y}\in A_{\mathbf{x}}}P(\mathbf{x},\mathbf{y})V(\mathbf{y})+\sum_{\mathbf{y}\in R_{\mathbf{x}}}P(\mathbf{x},\mathbf{y})V(\mathbf{y})\notag \\
&=&\sum_{\mathbf{y}\in A_{\mathbf{x}}}q(\mathbf{x},\mathbf{y})V(\mathbf{y})+\sum_{\mathbf{y}\in R_{\mathbf{x}}}q(\mathbf{x},\mathbf{y})\frac{\pi(\mathbf{y})}{\pi(\mathbf{x})}V(\mathbf{y})+\sum_{\mathbf{y}\in R_{\mathbf{x}}}q(\mathbf{x},\mathbf{y})\left[1-\frac{\pi(\mathbf{y})}{\pi(\mathbf{x})}\right]V(\mathbf{x}).\ \ \
\label{drift8}
\end{eqnarray}
\begin{eqnarray}
\frac{\sum_{\mathbf{y}\in\mathbb{Z}^n}P(\mathbf{x},\mathbf{y})V(\mathbf{y})}{V(\mathbf{x})}&=&\sum_{\mathbf{y}\in A_{\mathbf{x}}}q(\mathbf{x},\mathbf{y})\frac{\pi(\mathbf{x})^{1/2}}{\pi(\mathbf{y})^{1/2}}+\sum_{\mathbf{y}\in R_{\mathbf{x}}}q(\mathbf{x},\mathbf{y})\left[1-\frac{\pi(\mathbf{y})}{\pi(\mathbf{x})}+\frac{\pi(\mathbf{y})^{1/2}}{\pi(\mathbf{x})^{1/2}}\right].\ \ \ \ \ \ \ \
\label{drift9}
\end{eqnarray}
\vspace*{4pt}
\end{figure*}
\setcounter{equation}{\value{TempEqCnt}}

Equipped with minorisation and drift conditions, we are now in a position to prove the following theorem:

\begin{my2}
Given the invariant lattice Gaussian distribution $D_{\Lambda,\sigma,\mathbf{c}}$, the Markov chain established by the symmetric Metropolis-Klein algorithm is geometrically ergodic.
\end{my2}

\begin{proof}
First of all, the distribution $\pi(\mathbf{x})=D_{\Lambda,\sigma,\mathbf{c}}(\mathbf{x})$ is clearly bounded between $0$ and $1$ over any bounded set. Besides, for any $\|\mathbf{Bx}-\mathbf{By}\|\leq \delta_q$, where $\delta_q>0$ is a constant, the proposal distribution $q(\mathbf{x},\mathbf{y})$ can always be
lower bounded by a constant $\epsilon_q>0$ as follows,
\begin{eqnarray}
q(\mathbf{x},\mathbf{y})&\geq&\frac{e^{-\frac{\delta_q^2}{2\sigma^2}}}{\prod^n_{i=1}\rho_{\sigma_i, \widetilde{y}_i}(\mathbb{Z})}  \notag\\
&\overset{(j)}{\geq}&\frac{e^{-\frac{\delta_q^2}{2\sigma^2}}}{\prod^n_{i=1}\rho_{\sigma_i}(\mathbb{Z})}=\epsilon_q,
\end{eqnarray}
where $(j)$ holds due to (\ref{nnnnnn}). Thus, by \cite[Theorem 2.1]{RobertsGeometricconvergencea}, every non-empty bounded set $C\subseteq\mathbb{Z}^n$ in the underlying Markov chain of the SMK algorithm is a small set.
Then we may define a small set $C$ as
\begin{equation}
C=\{\mathbf{x}\in\mathbb{Z}^n:\pi(\mathbf{x})\geq\epsilon\}
\label{smalldefine}
\end{equation}
for sufficiently small $\epsilon$.

Meanwhile, at each Markov move, the acceptance ratio (\ref{MK-acceptance probability}) suggests the acceptance region $A_{\mathbf{x}}$ and the potential rejection
region $R_{\mathbf{x}}$ for current state $\mathbf{x}$ as follows:
\begin{eqnarray}
A_\mathbf{x}=\{\mathbf{y}\in\mathbb{Z}^n|\pi(\mathbf{y})\geq\pi(\mathbf{x})\}; \label{eq:A}\\
R_\mathbf{x}=\{\mathbf{y}\in\mathbb{Z}^n|\pi(\mathbf{y})<\pi(\mathbf{x})\}. \label{eq:R}
\end{eqnarray}
Obviously, state candidate $\mathbf{y}\in A_{\mathbf{x}}$ will surely be accepted by $\mathbf{X}_{t+1}$ while state candidate $\mathbf{y}\in R_{\mathbf{x}}$ has a certain probability to be rejected. Then, the LHS of the drift condition (\ref{drift5}) can be rewritten as (\ref{drift8}), where the second and third terms result from whether state candidate $\mathbf{y}\in R_{\mathbf{x}}$ is accepted or rejected, respectively.

Furthermore, setting the potential function $V(\mathbf{x})=\pi(\mathbf{x})^{-\frac{1}{2}}$. Dividing \eqref{drift8} by $V(\mathbf{x})$ on both sides, we have the expression shown in (\ref{drift9}).

\setcounter{equation}{79}
Since the ratios on the RHS of \eqref{drift9} $\leq 1$, we obtain\footnote{Note that $1\leq 1-a^2+a \leq \frac{5}{4}$ for $0\leq a \leq 1$.}
\begin{eqnarray}
\frac{\sum_{\mathbf{y}\in\mathbb{Z}^n}P(\mathbf{x},\mathbf{y})V(\mathbf{y})}{V(\mathbf{x})}&\leq&\frac{5}{4}.
\label{drift10}
\end{eqnarray}

\begin{figure}[t]
\begin{center}
\includegraphics[width=3.2in,height=2.0in]{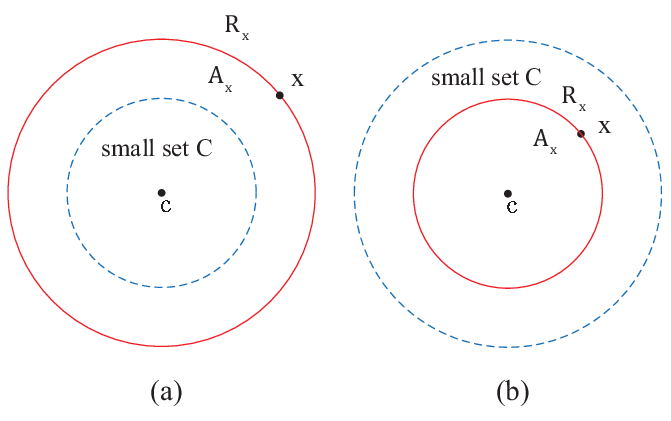}
\end{center}
\vspace{-1em}
  \caption{Illustration of cases (a) $\mathbf{x}\notin C$ and (b) $\mathbf{x}\in C$ in the Markov move induced by SMK. The blue dash circle represents the area of the small set while the red solid circle denotes the acceptance region $A_\mathbf{x}$.}
  \label{simulation 6}
\end{figure}

Depending on whether $\mathbf{x}\in C$ or not, the drift condition can be rewritten as:
\begin{equation}
\sum_{\mathbf{y}\in\Omega}P(\mathbf{x},\mathbf{y})V(\mathbf{y})\leq\lambda V(\mathbf{x}) \ \ \ \text{for}\ \mathbf{x}\notin C
\label{drift11}
\end{equation}
and
\begin{equation}
\sum_{\mathbf{y}\in\Omega}P(\mathbf{x},\mathbf{y})V(\mathbf{y})\leq\lambda V(\mathbf{x})+b \ \ \ \text{for}\ \mathbf{x}\in C.
\label{drift22}
\end{equation}
The two cases are illustrated in Fig. 6. We proceed case by case.

(i). In the case $\mathbf{x}\in C$,
\begin{eqnarray}
V(\mathbf{x})\leq \frac{1}{\sqrt{\epsilon}}.
\end{eqnarray}
By \eqref{drift10} we have
\begin{equation}
\sum_{\mathbf{y}\in \mathbb{Z}^n}P(\mathbf{x},\mathbf{y})V(\mathbf{y})\leq \frac{1}{\sqrt{\epsilon}}\cdot \frac{5}{4} = b \ \ \ \text{for}\ \mathbf{x}\in C
\end{equation}
and thus condition \eqref{drift22} is satisfied.

%

(ii). In the case $\mathbf{x}\notin C$, we consider
\begin{eqnarray}
\lambda&=&\limsup_{\|\mathbf{x}\| \to \infty} \frac{\underset{\mathbf{y}\in\mathbb{Z}^n}{\sum}\hspace{-.5em}P(\mathbf{x},\mathbf{y})V(\mathbf{y})}{V(\mathbf{x})}.
\end{eqnarray}
If $\lambda < 1$, then \eqref{drift11} is satisfied for sufficient small $\epsilon$.

It is easy to verify that
\begin{equation}
\underset{\|\mathbf{x}\|\rightarrow\infty}{{\lim}}\ell(\mathbf{x})\cdot\nabla \log \pi(\mathbf{x})=-\infty,
\end{equation}
{where $\ell(\mathbf{x})$ denotes the unit vector $\mathbf{x}/\|\mathbf{x}\|$ and $\nabla$ represents the gradient. This condition implies that for any $\gamma>0$, there exists $R>0$ such that for $\|\mathbf{x}\|\geq R$}
\begin{equation}
\frac{\pi(\mathbf{x}+a\cdot \ell(\mathbf{x}))}{\pi(\mathbf{x})}\leq e^{-a\cdot\gamma}, \quad a\geq 0.
\end{equation}
In other words, as $\|\mathbf{x}\|$ goes to infinity, the above ratio is at least exponentially decaying with a rate $\gamma$ tending to infinity.

Let $C_{\zeta} = \{ \mathbf{x} \in \mathbb{R}^n \ | \ \pi(\mathbf{x})=\zeta \}$. We define the radial $\mu$-zone around $C_{\pi(\mathbf{x})}$ as
\[
C_{\pi(\mathbf{x})}(\mu) = \{\mathbf{z} + s \cdot \ell(\mathbf{z}) \ | \ \mathbf{z} \in C_{\pi(\mathbf{x})}, -\mu \leq s \leq \mu\}.
\]
Denote by $B(\mathbf{x},K)$ a Euclidean ball of radius $K$, centered at $\mathbf{x}$. As in \cite{JarnerGeometricergodicity}, for arbitrary but fixed $\epsilon' > 0$, choose $K>0$ such that
\[
\sum_{\mathbf{y}\in \mathbb{Z}^n,\mathbf{By}\notin B(\mathbf{Bx},K)}q(\mathbf{x},\mathbf{y}) \leq \epsilon'.
\]
This can be assured by noting that
\begin{eqnarray}
q(\mathbf{x},\mathbf{y})&=&\frac{e^{-\frac{\|\mathbf{Bx}-\mathbf{By}\|^2}{2\sigma^2}}}{\prod^n_{i=1}\rho_{\sigma_i, \widetilde{y}_i}(\mathbb{Z})}  \notag\\
&\overset{(l)}{\leq}&\frac{e^{-\frac{\|\mathbf{Bx}-\mathbf{By}\|^2}{2\sigma^2}}}{\prod^n_{i=1}\rho_{\sigma_i,1/2}(\mathbb{Z})},
\end{eqnarray}
where $(l)$ is because $\rho_{\sigma_i, \widetilde{y}_i}(\mathbb{Z})$ has a minimum at $\widetilde{y}_i=1/2$ and then applying a tail bound of lattice Gaussian distribution \cite[Lemma 1.5]{Banaszczyk}.

From the fact that the Euclidean norms $\{\|\mathbf{Bx}\|, \ \mathbf{x}\in \mathbb{Z}^n \}$ of a lattice are discrete, it follows that for any $K>0$ there exists $\mu>0$ such that
\begin{eqnarray}
\limsup_{\|\mathbf{x}\| \to \infty}  \sum_{\mathbf{y}\in \mathbb{Z}^n \cap C_{\pi(\mathbf{x})}(\mu) \above 0pt \mathbf{By} \in B(\mathbf{Bx},K)}\hspace{-1.5em}q(\mathbf{x},\mathbf{y}) \leq \limsup_{\|\mathbf{x}\| \to \infty}  \sum_{\mathbf{y}\in \mathbb{Z}^n, \|\mathbf{By}\|=\|\mathbf{Bx}\|\above 0pt \mathbf{By} \in B(\mathbf{Bx},K)}\hspace{-2em}q(\mathbf{x},\mathbf{y}).
\end{eqnarray}


Define two regions $\underline{A}_\mathbf{x}=\{\mathbf{y}\in\mathbb{Z}^n|\pi(\mathbf{y})>\pi(\mathbf{x})\}$ and $\overline{R}_\mathbf{x}=\{\mathbf{y}\in\mathbb{Z}^n|\pi(\mathbf{y})\leq\pi(\mathbf{x})\}$, which are slightly different from \eqref{eq:A}, \eqref{eq:R}, i.e., $\underline{A}_\mathbf{x}$ does not include the boundary but $\overline{R}_\mathbf{x}$ does. As $\|\mathbf{x}\| \to \infty$, the ratios in \eqref{drift9} tend to 0 outside of any radial $\mu$-zone whose $\mathbf{y} \in B(\mathbf{Bx},K)$ for any $K$, and we arrive at
\begin{eqnarray}
\lambda\hspace{-.5em}&\leq&\hspace{-.5em}\limsup_{\|\mathbf{x}\| \to \infty}  \sum_{\mathbf{y}\in \overline{R}_{\mathbf{x}}}q(\mathbf{x},\mathbf{y})\notag\\
\hspace{-.5em}&=&\hspace{-.5em}  1- \liminf_{\|\mathbf{x}\| \to \infty} \sum_{\mathbf{y}\in \underline{A}_{\mathbf{x}}}q(\mathbf{x},\mathbf{y})\notag\\
&\overset{(k)}{<}&1
\label{drift2}
\end{eqnarray}
where inequality $(k)$ holds because
\begin{equation}
\liminf_{\|\mathbf{x}\| \to \infty} \sum_{\mathbf{y}\in \underline{A}_{\mathbf{x}}}q(\mathbf{x},\mathbf{y})>0
\end{equation}
due to symmetry of $q(\mathbf{x},\mathbf{y})$. In fact, as shown in Fig. \ref{simulation 6}, it follows from symmetry that
\begin{equation}
\sum_{\mathbf{y}\in \underline{A}_{\mathbf{x}}}q(\mathbf{x},\mathbf{y})<\frac{1}{2}<\sum_{\mathbf{y}\in \overline{R}_{\mathbf{x}}}q(\mathbf{x},\mathbf{y}),
\end{equation}
and the two probabilities can approach $\frac{1}{2}$ as $\|\mathbf{x}\| \to \infty$.
This completes the proof in the case $\mathbf{x}\notin C$.

\end{proof}

In essence, the convergence of geometric ergodicity can be classified into two stages. On one hand, if $\mathbf{x}\notin C$, the drift condition guarantees the Markov chain shrinks geometrically towards the small set $C$. On the other hand, if $\mathbf{x}\in C$, the {minorisation condition} shown in (\ref{smallset}) implies the Markov chain will converge to the stationary distribution exponentially fast. This can be demonstrated by using the coupling technique as in the previous section and $\delta$ is just the exponential decay coefficient, which depends on $C$. It was shown in \cite{RosenthalMinorization} that, for $C=\{\mathbf{x}: V(\mathbf{x})\leq d\}$ and $d>2b/(1-\lambda)$, Markov chains satisfying the drift condition will converge exponentially to the stationary distribution as follows
\begin{equation}
\|\hspace{-.1em}P^n\hspace{-.1em}(\mathbf{x}_0, \hspace{-.2em}\cdot)-\pi(\cdot)\hspace{-.1em}\|_{TV}\hspace{-.2em}\leq\hspace{-.2em}(1\hspace{-.2em}-\hspace{-.1em}\delta)^{rn}\hspace{-.3em}+\hspace{-.1em}\left(\hspace{-.2em}\frac{U^r}{\alpha^{1-r}}\hspace{-.2em}\right)^{\hspace{-.2em}n}\hspace{-.4em}\left(\hspace{-.3em}1\hspace{-.2em}+\hspace{-.2em}\frac{b}{1\hspace{-.2em}-\hspace{-.2em}\lambda}\hspace{-.2em}+\hspace{-.2em}V\hspace{-.2em}(\mathbf{x}_0)\hspace{-.4em}\right)\hspace{-.3em},
\label{ddddd}
\end{equation}
where $0<r<1$,
\begin{equation}
\alpha=\frac{1+d}{1+2b+\lambda d}\ \ \text{and}\ \ U=1+2(d+b).
\end{equation}

Clearly, there is a trade-off between these two convergence stages: a larger set $C$ indicates a smaller $\delta$ in the minorisation condition for $\mathbf{x}\in C$ but a faster shrink speed $\lambda$ towards $C$ for $\mathbf{x}\notin C$ (close to $1/2$ when $\|\mathbf{x}\|\rightarrow\infty$). However, the size of $C$, measured by $d$ here, is determined artificially, making both $\delta$ and $\lambda$ sensitive to a slight change of $d$. Moreover, a closed-form expression of $\lambda$ is difficult to get even for a specific $C$. Therefore, although geometric ergodicity can be achieved by the proposed SMK algorithm, it is difficult to obtain quantitative bounds on $\delta$ and $\lambda$.

Finally, (\ref{ddddd}) indicates that the convergence of the Markov chain arising from the SMK algorithm also highly depends on the starting state $\mathbf{x}_0$, which follows the definition of geometric ergodicity given in (\ref{geo-ergodic}). In theory, $\mathbf{x}_0$ could be any candidate from the state space but a poor choice may intensively increase the required mixing time. To this end, starting the Markov chain with $\mathbf{x}_0$ as close to the center of the distribution as possible would be a judicious choice. This is actually in accordance with the result shown in (\ref{ddddd}), implying the closest point to $\mathbf{c}$ is the optimal choice. As a simple solution, Babai's nearest plane algorithm is recommended here to output $\mathbf{x}_0$ \cite{Babai}.

\section{Conclusions}
In this paper, two MH-based algorithms were proposed to sample from lattice Gaussian distributions. As the proposal distribution in the MH algorithms can be set freely, an independent proposal distribution and a symmetric proposal distribution were exploited respectively for geometric convergence. In addition, it was proven that the Markov chain arising from the independent MHK algorithm is uniformly ergodic, leading to exponential convergence regardless of the starting state. We showed its convergence rate can be explicitly calculated via theta series, making the mixing time predictable. On the other hand, the proposed SMK algorithm was demonstrated to be geometrically ergodic, where the selection of the starting state matters. Due to its inherent symmetry, it not only converges exponentially fast, but also is simple to implement.

\section*{Acknowledgment}
The authors would like to thank Damien Stehl\'{e}, Guillaume Hanrot and Antonio Campello for fruitful discussions, and Conghui Li for careful reading of the paper.

\appendices

\section{Proof of Inequality in (\ref{spectral1})}
\label{proofl22}
\begin{proof}
To start with, let us recall the definition of \emph{conductance} (also known as \emph{bottleneck ratio}) in Markov chains \cite{mixingtimemarkovchain}.
\begin{my3}
The conductance $\Phi$ of a Markov chain is defined as
\begin{equation}
\Phi(S)=\min_{S\subseteq\Omega,\pi(S)\leq1/2}\frac{Q(S,S^c)}{\pi(S)},
\end{equation}
where subset $S^c$ stands for the complement set of $S$ (i.e., $S\bigcup S^c=\Omega$, $S\bigcap S^c=\emptyset$), and the edge measure $Q$ is defined by
\begin{equation}
Q(x,y)=\pi(x)P(x,y)
\end{equation}
and
\begin{equation}
Q(S,S^c)=\hspace{-.5em}\sum_{x\in S, y\in S^c}\hspace{-1em}Q(x,y).
\end{equation}
\end{my3}
It is this value $0<\Phi\leq1$ that has been used to bound the spectral gap $\gamma$ of Markov
chains. More precisely, in the independent MHK algorithm, we have
\begin{flalign}
\Phi&=\min_{S\subseteq\Omega,\pi(S)\leq1/2}\frac{\sum_{\mathbf{x}\in S, \mathbf{y}\in S^c}\pi(\mathbf{x})P(\mathbf{x},\mathbf{y})}{\pi(S)}\notag\\
&\overset{(m)}{\geq}\min_{S\subseteq\Omega,\pi(S)\leq1/2}\frac{\sum_{\mathbf{x}\in S, \mathbf{y}\in S^c}\pi(\mathbf{x})\cdot\delta\pi(\mathbf{y})}{\pi(S)}\notag\\
&=\min_{S\subseteq\Omega,\pi(S)\leq1/2}\frac{\delta\cdot\sum_{\mathbf{x}\in S}\pi(\mathbf{x})\cdot\sum_{\mathbf{y}\in S^c}\pi(\mathbf{y})}{\pi(S)}\notag\\
&=\min_{S\subseteq\Omega,\pi(S)\leq1/2}\delta\cdot\pi(S^c)\notag\\
&\geq\frac{\delta}{2},
\end{flalign}
where inequality $(m)$ holds due to (\ref{mmmmm}).

Next, by invoking the \emph{cheeger inequality} \cite{SinclairImprovedbounds} of Markov chains
\begin{equation}
\frac{\Phi^2}{2}\leq\gamma\leq2\Phi,
\end{equation}
we have
\begin{equation}
\gamma\geq\frac{\delta^2}{8},
\end{equation}
completing the proof.
\end{proof}

\section{Proof of Lemma 2}
\label{proofl2}

\begin{proof}
According to the QR-decomposition $\mathbf{B}=\mathbf{QR}$, we have
\begin{equation}
q(\mathbf{x},\mathbf{y})=\frac{e^{-\frac{1}{2\sigma^2}\|\mathbf{Bx}-\mathbf{By}\|^2}}{\prod^n_{i=1}\rho_{\sigma_i, \widetilde{y}_i}(\mathbb{Z})}=\frac{e^{-\frac{1}{2\sigma^2}\|\mathbf{Rx}-\mathbf{Ry}\|^2}}{\prod^n_{i=1}\rho_{\sigma_i, \widetilde{y}_i}(\mathbb{Z})}
\label{lllllll}
\end{equation}
by removing the orthogonal matrix $\mathbf{Q}$, where $\widetilde{y}_i=\frac{c'_i-\sum^n_{j=i+1}r_{i,j}y_j}{r_{i,i}}$, $\mathbf{c'}=\mathbf{Rx}$.

Specifically, the term $\rho_{\sigma_i, \widetilde{y}_i}(\mathbb{Z})$ in the denominator of (\ref{lllllll}) can be expressed as
\begin{eqnarray}
\rho_{\sigma_i, \widetilde{y}_i}(\mathbb{Z})&=&\sum_{z_i\in\mathbb{Z}}e^{-\frac{1}{2\sigma_i^2}(z_i-\frac{c'_i-\sum^n_{j=i+1}r_{i,j}y_j}{r_{i,i}})^2}\notag\\
&=&\sum_{z_i\in\mathbb{Z}}e^{-\frac{1}{2\sigma_i^2}(z_i-\frac{\sum^n_{j=i}r_{i,j}x_j-\sum^n_{j=i+1}r_{i,j}y_j}{r_{i,i}})^2}\notag\\
&=&\sum_{z_i\in\mathbb{Z}}e^{-\frac{1}{2\sigma_i^2}(x_i-z_i+\overset{n}{\underset{j=i+1}{\sum}}\frac{r_{i,j}}{r_{i,i}}(x_j-y_j))^2}\notag\\
&=&\sum_{z'_i\in\mathbb{Z}}e^{-\frac{1}{2\sigma_i^2}(z'_i-\phi)^2}\notag\\
&=&\rho_{\sigma_i, \phi}(\mathbb{Z}), \label{eq:107}
\end{eqnarray}
where $z'_i=z_i-x_i$ and $\phi=\overset{n}{\underset{j=i+1}{\sum}}\frac{r_{i,j}}{r_{i,i}}(x_j-y_j)$.

Similarly, we can easily get that
\begin{eqnarray}
\rho_{\sigma_i, \widetilde{x}_i}(\mathbb{Z})&=&\sum_{z_i\in\mathbb{Z}}e^{-\frac{1}{2\sigma_i^2}(y_i-z_i+\overset{n}{\underset{j=i+1}{\sum}}\frac{r_{i,j}}{r_{i,i}}(y_j-x_j))^2}\notag\\
&=&\sum_{z'_i\in\mathbb{Z}}e^{-\frac{1}{2\sigma_i^2}(z'_i-\phi)^2}\notag\\
&=&\rho_{\sigma_i, \phi}(\mathbb{Z})=\rho_{\sigma_i, \widetilde{y}_i}(\mathbb{Z}),
\end{eqnarray}
where $\widetilde{x}_i=\frac{c''_i-\sum^n_{j=i+1}r_{i,j}x_j}{r_{i,i}}$, $\mathbf{c''}=\mathbf{Ry}$. Therefore, we have $q(\mathbf{x},\mathbf{y})=q(\mathbf{y},\mathbf{x})$.

In fact, \eqref{eq:107} shows that $q(\mathbf{x},\mathbf{y})$ is a function of $\mathbf{x}-\mathbf{y}$ only; moreover, since $\rho_{\sigma_i, \phi}(\mathbb{Z})$ is even in $\phi$, $q(\mathbf{x},\mathbf{y})=q(\mathbf{x}-\mathbf{y})=q(\mathbf{y}-\mathbf{x})$, completing the proof.
\end{proof}

\bibliographystyle{IEEEtran}
\bibliography{IEEEabrv,reference1}

\begin{thebibliography}{10}
\providecommand{\url}[1]{#1}
\csname url@samestyle\endcsname
\providecommand{\newblock}{\relax}
\providecommand{\bibinfo}[2]{#2}
\providecommand{\BIBentrySTDinterwordspacing}{\spaceskip=0pt\relax}
\providecommand{\BIBentryALTinterwordstretchfactor}{4}
\providecommand{\BIBentryALTinterwordspacing}{\spaceskip=\fontdimen2\font plus
\BIBentryALTinterwordstretchfactor\fontdimen3\font minus
  \fontdimen4\font\relax}
\providecommand{\BIBforeignlanguage}[2]{{%
\expandafter\ifx\csname l@#1\endcsname\relax
\typeout{** WARNING: IEEEtran.bst: No hyphenation pattern has been}%
\typeout{** loaded for the language `#1'. Using the pattern for}%
\typeout{** the default language instead.}%
\else
\language=\csname l@#1\endcsname
\fi
#2}}
\providecommand{\BIBdecl}{\relax}
\BIBdecl

\bibitem{Banaszczyk}
W.~Banaszczyk, ``New bounds in some transference theorems in the geometry of
  numbers,'' \emph{Math. Ann.}, vol. 296, pp. 625--635, 1993.

\bibitem{Forney_89}
G.~Forney and L.-F. Wei, ``Multidimensional constellations--{Part II}: Voronoi
  constellations,'' \emph{IEEE J. Sel. Areas Commun.}, vol.~7, no.~6, pp.
  941--958, Aug. 1989.

\bibitem{Kschischang_Pasupathy}
F.~R. Kschischang and S.~Pasupathy, ``Optimal nonuniform signaling for
  {G}aussian channels,'' \emph{{IEEE} Trans. Inform. Theory}, vol.~39, pp.
  913--929, May. 1993.

\bibitem{LB_13}
C.~Ling and J.-C. Belfiore, ``Achieiving the {AWGN} channel capacity with
  lattice {Gaussian} coding,'' \emph{IEEE Trans. Inform. Theory}, vol.~60,
  no.~10, pp. 5918--5929, Oct. 2014.

\bibitem{LLBS_12}
C.~Ling, L.~Luzzi, J.-C. Belfiore, and D.~Stehl\'{e}, ``Semantically secure
  lattice codes for the {G}aussian wiretap channel,'' \emph{IEEE Trans. Inform.
  Theory}, vol.~60, no.~10, pp. 6399--6416, Oct. 2014.

\bibitem{7360779}
H.~Mirghasemi and J.~C. Belfiore, ``Lattice code design criterion for {MIMO}
  wiretap channels,'' in \emph{2015 IEEE Information Theory Workshop - Fall
  (ITW)}, Oct 2015, pp. 277--281.

\bibitem{7058433}
S.~Vatedka, N.~Kashyap, and A.~Thangaraj, ``Secure compute-and-forward in a
  bidirectional relay,'' \emph{IEEE Transactions on Information Theory},
  vol.~61, no.~5, pp. 2531--2556, May 2015.

\bibitem{MicciancioGaussian}
D.~Micciancio and O.~Regev, ``Worst-case to average-case reductions based on
  {Gaussian} measures,'' in \emph{Proc. Ann. Symp. Found. Computer Science},
  Rome, Italy, Oct. 2004, pp. 372--381.

\bibitem{GentryDissertation}
C.~Gentry, ``A fully homomorphic encryption scheme,'' Ph.D. dissertation,
  Stanford University, USA, 2009.

\bibitem{RegevSolvingtheShortestVectorProblem}
D.~Aggarwal, D.~Dadush, O.~Regev, and N.~Stephens-Davidowitz, ``Solving the
  shortest vector problem in $2^n$ time via discrete {Gaussian} sampling,''
  \emph{STOC}, 2015.

\bibitem{RegevSolvingtheClosestVectorProblem}
D.~Aggarwal, D.~Dadush, and N.~Stephens-Davidowitz, ``Solving the closest
  vector problem in $2^n$ time --- the discrete {Gaussian} strike again!''
  \emph{FOCS}, 2015.

\bibitem{CongRandom}
S.~Liu, C.~Ling, and D.~Stehl\'{e}, ``{Decoding by sampling: A randomized
  lattice algorithm for bounded distance decoding},'' \emph{IEEE Trans. Inform.
  Theory}, vol.~57, pp. 5933--5945, Sep. 2011.

\bibitem{DerandomizedJ}
Z.~Wang, S.~Liu, and C.~Ling, ``Decoding by sampling - {Part} {II}:
  Derandomization and soft-output decoding,'' \emph{IEEE Trans. Commun.},
  vol.~61, no.~11, pp. 4630--4639, Nov. 2013.

\bibitem{DGStoCVPSVP}
N.~Stephens-Davidowitz, ``Discrete {Gaussian} sampling reduces to {CVP} and
  {SVP},'' submitted for publication. [Online]. Available:
  http://arxiv.org/abs/1506.07490.

\bibitem{Klein}
P.~Klein, ``Finding the closest lattice vector when it is unusually close,'' in
  \emph{ACM-SIAM Symp. Discr. Algorithms}, 2000, pp. 937--941.

\bibitem{Trapdoor}
C.~Gentry, C.~Peikert, and V.~Vaikuntanathan, ``Trapdoors for hard lattices and
  new cryptographic constructions,'' in \emph{Proc. 40th Ann. ACM Symp. Theory
  of Comput.}, Victoria, Canada, 2008, pp. 197--206.

\bibitem{ZhengWangMCMCLatticeGaussian}
Z.~Wang, C.~Ling, and G.~Hanrot, ``{Markov chain Monte Carlo} algorithms for
  lattice {Gaussian} sampling,'' in \emph{Proc. IEEE International Symposium on
  Information Theory (ISIT)}, Honolulu, USA, Jun. 2014, pp. 1489--1493.

\bibitem{HassibiMCMCnew}
B.~Hassibi, M.~Hansen, A.~Dimakis, H.~Alshamary, and W.~Xu, ``Optimized {Markov
  Chain Monte Carlo} for signal detection in {MIMO} systems: {An} analysis of
  the stationary distribution and mixing time,'' \emph{IEEE Transactions on
  Signal Processing,}, vol.~62, no.~17, pp. 4436--4450, Sep. 2014.

\bibitem{McmcDatta}
T.~Datta, N.~Kumar, A.~Chockalingam, and B.~Rajan, ``A novel {Monte Carlo}
  sampling based receiver for large-scale uplink multiuser {MIMO} systems,''
  \emph{IEEE Transactions on Vehicular Technology,}, vol.~62, no.~7, pp.
  3019--3038, Sep. 2013.

\bibitem{MCMCBehrouz}
B.~Farhang-Boroujeny, H.~Zhu, and Z.~Shi, ``Markov chain {Monte Carlo}
  algorithms for {CDMA} and {MIMO} communication systems,'' \emph{IEEE Trans.
  Signal Process.}, vol.~54, no.~5, pp. 1896--1909, 2006.

\bibitem{ChenMCMC}
R.~Chen, J.~Liu, and X.~Wang, ``Convergence analyses and comparisons of
  {Markov} {chain} {Monte Carlo} algorithms in digital communications,''
  \emph{IEEE Trans. on Signal Process.}, vol.~50, no.~2, pp. 255--270, 2002.

\bibitem{XiaodongWangMultilevel}
P.~Aggarwal and X.~Wang, ``Multilevel sequential {M}onte {C}arlo algorithms for
  {MIMO} demodulation,'' \emph{IEEE Transactions on Wireless Communications},
  vol.~6, no.~2, pp. 750--758, Feb. 2007.

\bibitem{MCMCHaidongZhu}
H.~Zhu, B.~Farhang-Boroujeny, and R.-R. Chen, ``On performance of sphere
  decoding and {Markov} chain {Monte} {Carlo} detection methods,'' \emph{IEEE
  Signal Processing Letters}, vol.~12, no.~10, pp. 669--672, 2005.

\bibitem{Hastings1970}
W.~K. Hastings, ``{Monte Carlo} sampling methods using {Markov} chains and
  their applications,'' \emph{Biometrika}, vol.~57, pp. 97--109, 1970.

\bibitem{WangxiaodongMCMag}
A.~Doucet and X.~Wang, ``Monte {Carlo} methods for signal processing,''
  \emph{IEEE Signal Process. Mag.}, vol.~22, no.~6, pp. 152 -- 170, Nov. 2005.

\bibitem{RobertsGeneralstatespace}
G.~O. Roberts, ``General state space {Markov} chains and {MCMC} algorithms,''
  \emph{Probability Surveys}, vol.~1, pp. 20--71, 2004.

\bibitem{LLLoriginal}
A.~K. Lenstra, H.~W. Lenstra, and L.~Lovasz, ``Factoring polynomials with
  rational coefficients,'' \emph{Math. Annalen}, vol. 261, pp. 515--534, 1982.

\bibitem{mixingtimemarkovchain}
D.~A. Levin, Y.~Peres, and E.~L. Wilmer, \emph{Markov Chains and Mixing Time},
  American Mathematical Society, 2008.

\bibitem{RosenthalOPDAMCMC}
J.~S. Rosenthal, ``Optimal proposal distributions and adaptive {MCMC},''
  \emph{Handbook of Markov chain Monte Carlo: Methods and Applications.},
  Brooks, S.P., Gelman, A., Jones, G., Meng, X.-L. (eds.) Chapman and Hall/CRC
  Press, Florida, USA. 2010.

\bibitem{RapidlyMixingRandall}
D.~Randall, ``Rapidly mixing {Markov} chains with applications in computer
  science and physics,'' \emph{Computing in Science and Engineering}, vol.~8,
  no.~2, pp. 30--41, 2006.

\bibitem{Meynbook}
S.~P. Meyn and R.~L. Tweedie, \emph{Markov Chains and Stochastic
  Stability}.\hskip 1em plus 0.5em minus 0.4em\relax UK, Cambridge University
  Press, 2009.

\bibitem{JarnerPolynomialconvergence}
S.~F. Jarner and G.~O. Roberts, ``Polynomial convergence rates of {Markov}
  chains,'' \emph{Ann. Appl. Probab.}, vol.~12, pp. 224--247, 2002.

\bibitem{MetropolisOrignial}
N.~Metropolis, A.~W. Rosenbluth, M.~N. Rosenbluth, A.~H. Teller, and E.~Teller,
  ``{Equations of state calculations by fast computing machines},'' \emph{J.
  Chem. Phys.}, vol.~21, pp. 1087--1091, 1953.

\bibitem{MengersenRate}
K.~L. Mengersen and R.~L. Tweedie, ``Rates of convergence of the {Hastings} and
  {Metropolis} algorithms,'' \emph{Ann. Statist.}, vol.~24, pp. 101--121, 1996.

\bibitem{ConwayandSloane}
J.~H. Conway and N.~A. Sloane, \emph{Sphere Packings, Lattices and
  Groups}.\hskip 1em plus 0.5em minus 0.4em\relax New York: Springer-Verlag,
  1998.

\bibitem{BelfioreConstructionandAnalysis}
F.~Oggier, P.~Sol\'{e}, and J.-C. Belfiore, ``Lattice codes for the wiretap
  {Gaussian} channel: Construction and analysis,'' \emph{IEEE Trans. Inform.
  Theory.}, accepted, 2015.

\bibitem{CongProxity}
C.~Ling, ``On the proximity factors of lattice reduction-aided decoding,''
  \emph{IEEE Trans. Signal Process.}, vol.~59, no.~6, pp. 2795--2808, Jun.
  2011.

\bibitem{ThetaExplicit}
J.~Yi, ``Theta-function identities and the explicit formulas for
  {T}heta-function and their applications,'' in \emph{Proc J. Math. Anal.
  Appl.}, vol. 292, 2004, pp. 381--400.

\bibitem{ThetaBook}
B.~Berndt, \emph{Ramanujan's Notebooks, Part V}.\hskip 1em plus 0.5em minus
  0.4em\relax Springer-Verlag, New York, 1998.

\bibitem{RobertsGeometricconvergencea}
G.~O. Roberts and R.~L. Tweedie, ``Geometric convergence and central limit
  theorems for multidimensional {Hastings} and {Metropolis} algorithms,''
  \emph{Biometrika}, vol.~83, pp. 95--110, 1996.

\bibitem{JarnerGeometricergodicity}
S.~F. Jarner and E.~Hansen, ``Geometric ergodicity of {Metropolis} algorithm,''
  \emph{Stochastic Process}, vol.~85, pp. 341--361, 2000.

\bibitem{RosenthalMinorization}
J.~S. Rosenthal, ``Minorization conditions and convergence rates for {Markov}
  chain {Monte} {Carlo},'' \emph{J. Amer. Statist. Assoc.}, vol.~90, pp.
  558--566, 1995.

\bibitem{Babai}
L.~Babai, ``On {L}ov\'asz' lattice reduction and the nearest lattice point
  problem,'' \emph{Combinatorica}, vol.~6, no.~1, pp. 1--13, 1986.

\bibitem{SinclairImprovedbounds}
A.~J. Sinclair, ``Improved bounds for mixing rates of {Markov} chains and
  multicommodity flow,'' \emph{Combin. Probab. Comput.}, vol.~1, pp. 351--370,
  1992.

\end{thebibliography}

\end{document}